\documentclass[11pt]{article}

\usepackage{amsfonts,amssymb,amsmath,latexsym,enumitem,ae,aecompl}
\usepackage{graphicx}

\usepackage{amsfonts,amssymb,amsmath,latexsym,bbm,ae,aecompl}
\usepackage{algorithm,algpseudocode,color,bbm}

\newcommand{\FF}{\vspace*{\medskipamount}}

\newcommand{\BBB}{\vspace*{-\bigskipamount}}

\newcommand{\cI}{{\mathcal I}}

\newcommand{\cN}{{\mathcal N}}
\newcommand{\cO}{\mathcal{O}}
\newcommand{\cP}{{\mathcal P}}

\newcommand{\cS}{{\mathcal S}}
\newcommand{\cT}{{\mathcal T}}

\newcommand{\dist}{\text{dist}}
\newcommand{\mE}{\mathbb{E}\,}

\newcommand{\Paragraph}[1]{\BBB\paragraph{#1}}

\newcommand{\qed}{\hfill $\square$ \smallskip}
\newcommand{\polylog}{\mbox{ polylog }}

\newenvironment{proof}{\noindent{\bf Proof:}}{\qed}

\newtheorem{theorem}{Theorem}
\newtheorem{lemma}{Lemma}
\newtheorem{fact}{Fact}

\newtheorem{proposition}{Proposition}

\newlength{\pagewidth}
\begin{document}

\baselineskip 	3ex
\parskip 		1ex

\title{Distributed Bare-Bones Communication in Wireless Networks \footnotemark[1]~\vfill}

\author{	Bogdan S. Chlebus~\footnotemark[2] \and
		Dariusz R. Kowalski~\footnotemark[3] \and
		Shailesh Vaya~\footnotemark[4]}

\footnotetext[1]{A preliminary version of this paper appeared as~\cite{ChlebusV16}.}

\footnotetext[2]{School of Computer and Cyber Sciences, 
			Augusta University, 
			Augusta, Georgia, USA. 
			Work partly supported by the NSF Grant 1016847.}
			
\footnotetext[3]{School of Computer and Cyber Sciences, 
			Augusta University, 
			Augusta, Georgia, USA. }

\footnotetext[4]{Xerox Research Centre India, Bangalore, India.}

\date{}

\maketitle

\vfill


\begin{abstract}
We consider wireless networks operating under the SINR model of interference.
Nodes have limited individual knowledge and capabilities: 
they do not know their positions in a coordinate system  in the plane, 
further they do not know their neighborhoods, 
nor do they know the size of the network $n$, 
and finally they  cannot sense collisions resulting from simultaneous transmissions by
at least two neighbors.
Each node is equipped with a unique integer name, where $N$ as an upper bound on the a range of names. 
We refer  as a  backbone to a subnetwork induced by a diameter-preserving dominating set of nodes. 
Let  $\Delta$ denote a maximum number of nodes that can successfully receive a message transmitted by a node when no other nodes transmit concurrently.
We study distributed algorithms for  communication problems in three settings. 
In the single-node-start case, when one node starts an execution and other nodes are awoken by receiving messages from already awoken nodes, we present a randomized broadcast algorithm that wakes up all nodes in  $\cO(n \log^2 N)$ rounds with high probability. 
For the synchronized-start case, when all nodes start an execution simultaneously, we give a randomized algorithm  computing a backbone in $\cO(\Delta\log^{7} N)$ rounds with high probability.
In the partly-coordinated-start case, when a number of nodes start an execution together and other nodes are awoken by receiving messages from the already awoken nodes, we develop an algorithm that creates a backbone in time $\cO(n\log^2 N +\Delta\log^{7} N)$ with high probability.

\vfill

\end{abstract}

\vfill

\thispagestyle{empty}

\setcounter{page}{0}

\newpage

\section{Introduction}
\label{intro}

We consider wireless networks in which the effects of interference are determined by the signal-to-interference-plus-noise ratio (SINR) model. 
The extent to which such networks can support distributed communication depends on nodes' capabilities, like the ability to detect signals' collisions (caused by two or more neighbors transmitting simultaneously), and on the information that can be used in codes of algorithms, such as coordinates of nodes in a Cartesian coordinate system in a plane.
We demonstrate that efficient distributed communication can be carried out by a wireless network whose nodes have severely limited power.

We assume that the nodes of a wireless network are positioned in a two-dimensional Euclidean space.
This is abstracted into an associated graph structure called the communication graph of the network.
The  nodes of the network serve as vertices of the communication graph. 
A pair of such vertices $(u,w)$  makes an edge in the communication graph if~$w$ can successfully receive a message transmitted by~$u$ when no other node transmits simultaneously, and vice versa. 
We use $\Delta$ to denote a maximum node-degree in a communication graph, and $D$ to denote a  diameter of this graph.

We want to distinguish certain vertices of a communication graph such that they together make a \emph{backbone} of the graph.
We call these vertices  \emph{backbone vertices}. 
Backbone vertices induce a subgraph of a constant degree in the communication graph that has its diameter asymptotically equal to that of the whole communication graph and such that every node outside the backbone is connected to some backbone vertex.

Nodes of a wireless network communicate directly subject to restrictions on the signal-to-interference-plus-noise ratio (SINR). 
A transmission is successfully received by a node depending on a ratio of the signal strength to the strength of other signals plus ambient noise, when evaluated at a receiver.
Let $\cT$ be a set of nodes that transmit together at a round, and let two nodes $v$ and~$u$ be such that $v \in \cT$ and $u \notin \cT$.
The signal strength of $v$'s transmission as reaching~$u$ is determined as  $\cP(v,u) = P_v\cdot\dist(v,u)^{-\alpha}$, where $\dist(v,u)$ is the distance between $u$ and $v$, quantity~$p_v$ is the transmission strength, and $\alpha> 2$ is a path loss.
The interference at~$u$ means $\cI(v,u,\cT)=\sum_{w\in\cT\setminus\{v\}} \cP(w,u)$. 
The quantity SINR\,$(v,u,\cT)$  is defined as $\textrm{SINR}(v,u,\cT)=
\cP(v,u) / (\cN+\cI(v,u,\cT))$, where $\cN > 0$ is the ambient noise.
For a node~$u$ to successfully receive a transmission by node~$v$, it is necessary for the inequality $\textrm{SINR}(v,u,\cT) \ge \beta$ to hold, where $\beta>0$ is some threshold.
We work with a SINR model that combines weak-connectivity assumptions, as formulated by Daum et al.~\cite{DGKN13}, with weak-sensitivity (weak devices) assumptions, as formulated by Jurdzi\'nski et al.~\cite{JKS-ICALP-13}, see also~Jurdzi\'nski and Kowalski~\cite{JurdzinskiK-encyclopedia}.
In a weak-sensitivity setting, a node can never successfully receive  a transmission from the borderline of its transmission range, while weak connectivity means that all links that could successfully transfer a message in a suitably favorable scenario are included as links of a network,  see Section~\ref{sec:technical-preliminaries} for details.
We also assume that nodes cannot  detect collisions produced by interfering transmissions from at least two neighbors.

Algorithms are restricted in what information can be used in their codes.
A network is said to be \emph{bare-bones} when it is subject to a specific set of such restrictions, which are used in this paper and can be summarized as follows.
Each node among the $n$ nodes in a network has a unique name in the range $\{1, \dots, N\}$, for some positive integer $N\ge n$.
Each node knows its name and the number~$N$.
Nodes do not know their neighbors in the communication graph but they know the parameter~$\Delta$.

We consider distributed algorithms to synchronize and organize a wireless network.
The way a communication task is initiated impacts how a communication algorithm may  be designed.
We consider the following natural modes of initialization of communication tasks.
If all the nodes begin an execution together in the same round, then such an execution is said to be performed from a \emph{synchronized start}.
Executing algorithms from an \emph{single-node start} means that just a single node is awoken at the beginning of an execution, while all the other nodes do not send messages until they hear a message.
A case of \emph{multiple-nodes synchronized start} means that a group of nodes start an execution together, while the remaining nodes wait to be awoken by hearing messages.
Synchronized start is a global configuration of a network that facilitates executing a communication algorithm with all nodes starting at the same time.

We consider implementations of three communication primitives useful in developing  distributed algorithms.
One primitive is to prepare a synchronized-start configuration from a single-node start.
This means to make all nodes reach a ``start'' global state from which to begin  in unison an execution of some algorithm.
Another primitive is to synchronize start from a multiple-nodes synchronized start, which has a similar goal but  subject to the stated different initial conditions.
These two tasks can be accomplished by a broadcast with a wake-up functionality, like coordinating round numbers at nodes.
The third primitive is to build a backbone of the communication graph of a network, provided that a synchronized-start configuration has already been reached.

Next, we briefly summarize the results presented in this paper, and put them in the context of previous and related work.

\Paragraph{A summary of the results.}

We develop randomized distributed algorithms implementing certain communication primitives in wireless networks operating under the SINR regime of interferences among concurrent transmissions.
Each of them has the property that  a node uses only $\cO(\log^3 N)$ random bits in an execution.
The model is of weak-sensitivity and weak-connectivity.

We begin by developing a randomized distributed algorithm to perform a broadcast from a single-node start in $\cO(n \log^2 N)$ expected time, where it is the source that starts as an activated node. 
Once an execution is completed, every node will have been activated and all the nodes have  their round numbers synchronized.
This may be used as a preparation to begin an execution of a distributed communication algorithm from a synchronized start.

We present two randomized algorithms to build a backbone.
One of them completes the task from a synchronized start in  $\cO(\Delta\log^{7} N)$ rounds with high probability.
The other one creates a backbone from multiple-nodes coordinated-start in $\cO(n\log^2 N +\Delta\log^{7} N)$ rounds with high probability.

There are known lower bounds $\Omega(n\log N)$ and $\Omega(D\Delta)$ on time to broadcast from a single-node start, given by Jurdzi\'nski et al.~\cite{JKS-ICALP-13}.
These lower bounds hold for randomized algorithms and when nodes know their coordinates in a system of coordinates, so these lower bounds apply to our less demanding settings as well.
Therefore the time performance of our broadcast  algorithm misses a lower bound $\Omega(n\log N)$ by a factor of~$\cO(\log n)$.

Building a backbone from a single-node start could proceed by way of first coordinating all the nodes such that they can start simultaneously an execution of a distributed communication algorithm to build a backbone.
For $D=\Delta=\Theta(\sqrt{n})$, both constructing a backbone and performing broadcast could be performed in $\cO(\Delta \polylog N) = \cO(\sqrt{n} \polylog N)$ rounds, while they may need as much as $\Omega(D\Delta)=\Omega(n)$ rounds  for a single-node start. 
The performance of our algorithm for building a backbone from a synchronized start implies that having all the nodes of a network synchronized, so that they can start an execution
simultaneously, makes it possible to perform some distributed tasks faster than otherwise, in that  building a virtual backbone network is among such tasks.

Our work  extends the results of Jurdzi\'nski and Kowalski~\cite{JK-DISC12} and Jurdzi\'nski et al.~\cite{JKS-ICALP-13}, whose solutions rely on the knowledge of coordinates of nodes in a coordinate system in a plane, to a setting where such knowledge is not available.
These are the first algorithms with performance bounds close to optimal in the  model of weak sensitivity and weak connectivity  of wireless networks, in view of the lower bounds in the settings considered in Jurdzi\'nski et al~\cite{JKS-ICALP-13} that apply to the setting of this paper as well.

We introduce a novel approach to collision avoidance, based on strongly selective families with specifically chosen parameters, in order to compensate for lack of node coordinates in a system of coordinates in the plane.
We combine them with a number of graph-related and geometry-related algorithmic techniques to synchronize and build a backbone sub-network.

\Paragraph{Previous work.}

Communication algorithms in SINR wireless networks known in the literature have been designed under various assumptions regarding the underlying models. 
Such specifications are sometimes mutually exclusive so that the respective algorithms cannot be directly compared.  
To put our work in a proper context, we first clarify the relevant aspects of wireless networks and then concentrate on the broadcasting primitive.

We categorize communication models of SINR wireless networks following a methodology popularized by Jurdzi\'nski and Kowalski~\cite{JurdzinskiK-encyclopedia}.
This methodology is based to the following two independent criteria.
One pertains to weak versus strong ``sensitivity,'' which is also known as weak versus strong ``devices,'' according to the terminology used by Jurdzi\'nski et al.~\cite{JKS-ICALP-13}.
The other relates to weak versus strong ``links'' in reachability graphs, according to the terminology introduced by Daum et al.~\cite{DGKN13}.
These assumptions together determine four different settings, which we explain in detail in Section~\ref{sec:technical-preliminaries}.
Intuitively, weak-sensitivity makes it impossible for a node to ever receive a transmission from the borderline of its transmission range, while strong sensitivity determines a success of a transmission entirely by a suitable magnitude of the SINR ratio.
Weak connectivity means that all links that could possibly transfer a message under favorable circumstances are considered as valid links, unlike strong connectivity, in which borderline neighbors are not connected by links.
Along with these stipulations, additional assumptions determine which parameters of a network are known to the nodes, so that they can be used in codes of algorithms.


\newcommand{\RB}{\raisebox{12pt}{}}
\newcommand{\LB}{\raisebox{-6pt}{}}

\begin{table}[t]
\begin{center}
\begin{tabular}{|c| c || l | l |}
\hline
\RB
&& strong  connectivity: & weak connectivity: \\
\LB
& bounds &   $\varepsilon_c>\varepsilon_s$ & $\varepsilon_c=\varepsilon_s$  \\
\hline
\hline
\RB \LB
strong sensitivity: & lower &
$\Omega(D)$ $^\dagger$ & $\Omega(n)$ \cite{DGKN13}\\
\cline{2-4}
\RB \LB 
$\varepsilon_s=0$ & upper &  $\cO(D\log n\log^{\alpha+1}R_s)$ \cite{DGKN13}; $\cO(D\log^2n)$ \cite{JurdzinskiKRS-PODC14} & $\cO(n\log^2 n)$ $^\ddagger$ \cite{DGKN13}\\
\hline
\hline
\RB \LB 
weak sensitivity: & lower & $\Omega(D)$ $^\dagger$
& $\Omega(\min\{D\Delta,n\})$ \cite{JKS-ICALP-13} \\
\cline{2-4}
\RB \LB 
$\varepsilon_s>0$ & upper & $\cO(D\log^2n)$ $^\ast$ \cite{YuHWTL-SIROCCO12} & $\cO(n\log^2 n)$ this paper\\
\hline
\end{tabular}

\caption{\label{tab:1} 
Time performance bounds on randomized broadcasting for the four sensitivity and connectivity settings. 
The following parameters occur in the bounds: $\Delta$ denotes the maximum node degree in a communication graph, $D$ is  the diameter of this graph, $\alpha$ is the path loss, and $R_s$ is the maximum ratio between distances of neighbors in communication graph.
It is assumed that nodes' names are in a range $[1,N]$ such that $N=\cO(n)$ and $N$ is known to all nodes.
The lower bounds marked with the dagger~$\dagger$ follow by a distance argument.
The algorithm giving an upper bound marked with the asterisk~$\ast$ requires a power control mechanism and assumes $\varepsilon_c=2/3$.
The algorithm giving an upper bound marked with $\ddagger$ requires $\Omega(n\log^3 N)$  random bits per node.} 
\end{center}
\end{table}


\begin{table}[t]
\begin{center}

\begin{tabular}{|c| c || l | l |}
\hline
\RB
&& strong  connectivity: & weak connectivity: \\
\LB
& bounds &   $\varepsilon_c>\varepsilon_s$ & $\varepsilon_c=\varepsilon_s$  \\
\hline
\hline
\RB \LB
strong sensitivity: & lower &
$\Omega(\log n)$ $^\dagger$ \cite{KushilevitzM-SICOMP98} & $\Omega(\Delta)$ $^\ddagger$ \cite{JK-DISC12} \\
\cline{2-4}
\RB \LB 
$\varepsilon_s=0$ & upper &  $\widetilde{\cO}(\log^2 n)$ \cite{HalldorssonT21} & $\cO(N\Delta)$ $^\diamond$\\
\hline
\hline
\RB \LB 
weak sensitivity: & lower & $\Omega(\log n)$ $^\dagger$ \cite{KushilevitzM-SICOMP98}
& $\Omega(\Delta)$ $^\ddagger$ \cite{JK-DISC12} \\
\cline{2-4}
\RB \LB 
$\varepsilon_s>0$ & upper & $\cO(\log n)$ $^\ast$ \cite{YuZ0LY0C019,ZouYWYWH019} & $\cO(\Delta\log^7 n)$ this paper\\
\hline
\end{tabular}

\caption{\label{tab:2} 
Time performance bounds to build a backbone in a randomized manner for four sensitivity and connectivity settings, with spontaneous wake-up. 
The lower bounds marked with the dagger~$\dagger$ hold in the model of radio networks, which represents  interference through neighborhoods of nodes in graphs of arbitrary topology.
The upper bound marked with the asterisk~$\ast$ requires knowing the coordinates in the plane and the ability to exchange them in messages and assumes that the parameter~$\varepsilon_c$ is greater than some absolute positive constant.
The lower bounds marked with $\ddagger$ holds even with known node coordinates.
The algorithm giving the upper bound marked with the diamond $\diamond$ consists of executing single-node transmissions in a round-robin manner $\cO(\Delta)$ times.}
\end{center}
\end{table}

The  \emph{weak-sensitivity and weak-connectivity} model, which is used in this work, was considered by Jurdzi\'nski et al.~\cite{JKS-ICALP-13}.
They proved the lower bounds $\Omega(n\log N)$ and $\Omega(D\Delta)$ on time to broadcast from a single-node start.
These lower bounds hold for randomized algorithms even when the nodes know their coordinates in the plane.
They also developed a deterministic algorithm that accomplishes broadcast in time $\cO(\min\{D\Delta \log^2 N, n\log N\})$.
For this model, Jurdzi\'nski and Kowalski~\cite{JK-DISC12} gave a deterministic distributed algorithms building a backbone   from a synchronized start in $\cO(\Delta \polylog n)$ rounds.
The algorithms presented in Jurdzi\'nski and Kowalski~\cite{JK-DISC12} and Jurdzi\'nski  et al.~\cite{JKS-ICALP-13} rely on nodes knowing their position in a coordinate system.

For the model of \emph{weak sensitivity and strong connectivity}, Yu et al.~\cite{YuHWTL-SIROCCO12} gave broadcast algorithms operating in times $\cO(D+\log^2 n)$ and $\cO(D\log^2 n)$ with high probability, where the bound depends on how broadcast is initiated.
These algorithms additionally resort to a power-control mechanism.
This approach applies to broadcasting multiple messages and was generalized by Yu et al.~\cite{YuHWYL-INFOCOM13} to a scenario when nodes are activated in arbitrary rounds.
The model of wireless networks used in the papers~Yu et al.~\cite{YuHWTL-SIROCCO12} and Yu et al.~\cite{YuHWYL-INFOCOM13} incorporates additional parameters, with the suitable  assumptions on these parameters  on which efficiency of the algorithms depends;  one could expect that in such environments even more involved  communication tasks might have solutions with running time proportional to~$D$, with other parameters contributing sub-linear factors.

For the model of \emph{strong sensitivity and weak connectivity}, Daum et al.~\cite{DGKN13} gave a randomized broadcast algorithm operating in time $\cO(n\log^2 n)$.
Compared to the algorithm in this paper, the algorithm in~\cite{DGKN13} uses nearly exponentially more, namely $\Omega(n\log^3 N)$, random bits per node.
They showed a lower bound $\Omega(n)$, which holds in networks of diameter~$2$.

Finally, for the model of \emph{strong sensitivity and strong connectivity}, Jurdzi\'nski et al.~\cite{JKRS-DISC-13} gave a broadcast algorithm working in time $\cO(D\log n + \log^2 n)$ with high probability, which relies on nodes knowing their coordinates.
Jurdzi\'nski et al.~\cite{JurdzinskiKRS-PODC14} gave another algorithm that works in time $\cO(D \log^2 n)$ with high probability. 
The latter algorithm does not rely on nodes knowing their  coordinates, improving  the performance of algorithms for this model given by Daum et al.~\cite{DGKN13} for a suitable range of model parameters.
A solution in Jurdzi\'nski et al.~\cite{JurdzinskiKRS-PODC14} was  generalized to the wake-up problem with non-synchronized start-ups by Jurdzi\'nski et al.~\cite{JurdzinskiKRS-INFOCOM15}.
For this model of strong sensitivity and strong connectivity, Jurdzi\'nski et al.~\cite{JKS-FCT-13} studied deterministic solutions for the single-broadcast problem  when nodes know their own coordinates in the plane and those of their neighbors.
Their deterministic algorithm for a single-node start operates in  time $\cO(D \log^2 n)$, and another deterministic algorithm for a synchronized start operates in time $\cO(D + \log^2 n)$.
No deterministic algorithms for this model  are known that do not rely on the knowledge of coordinates of nodes in a coordinate system in the plane and that are of comparable time performance.

We summarize known lower and upper bounds on time performance of randomized broadcasting in Table~\ref{tab:1} and building backbone in Table~\ref{tab:2}.

Broadcasting can be considered with a limited goal to inform only the neighbors in the communication graph.
Such \emph{local broadcasting} in wireless networks was studied by Barenboim and Peleg~\cite{BarenboimP15},  Goussevskaia et al.~\cite{GoussevskaiaMW08}, Kesselheim and V{\"o}cking~\cite{KV10}, Yu et al.~\cite{YuWHL11},  Halld{\'o}rsson et al.~\cite{HalldorssonHL15}, Halld{\'o}rsson and Mitra~\cite{HalldorssonM12} and  Fuchs and Wagner~\cite{FuchsW13}.
Halld{\'o}rsson et al.~\cite{HalldorssonTWY16} studied broadcasting in dynamic wireless networks.
Centralized algorithms for the SINR model were surveyed by Goussevskaia et al.~\cite{GoussevskaiaPW10-survey}. 

Broadcasting of multiple messages was considered by Reddy et al.~\cite{ReddyKV15,ReddyV16} and Yu et al.~\cite{YuHWTL-SIROCCO12,YuHWYL-INFOCOM13}.
Chlebus et al.~\cite{ChlebusKPR-ICALP11} considered multi-communication primitives for radio networks; that paper also used breadth-then-depth trees, similarly as the broadcast algorithm in Section~\ref{sec:single-node-start}.
Derbel and Talbi~\cite{DerbelT10} showed how to estimate node degrees in radio networks with nodes initially not knowing their neighbors.

Scheideler et al.~\cite{ScheidelerRS08} gave an algorithm to find a dominating set in time $\cO(\log n)$.
That algorithm relies on manipulating thresholds in physical carrier sensing.
Halld{\'{o}}rsson and Tonoyan~\cite{HalldorssonT21} demonstrated how nodes in wireless networks can leverage indirect information received during collisions to infer neighborhoods  and construct backbones; building on this, they showed applications of backbones to various communication problems in wireless networks. 
Moses and Vaya~\cite{MosesV21} developed deterministic algorithms for multi-broadcast and building backbones in wireless networks.
Yu et al.~\cite{YuZ0LY0C019} considered distributed algorithms to construct dominating sets in dynamic wireless networks.
Zou et al.~\cite{ZouYWYWH019} studied building backbones in wireless networks subject to adversarial jamming.
Constructions and applications of virtual backbones and dominating sets in wireless networks were surveyed by Yu et al.~\cite{YuWWY2013}.

There has been a great variety of approaches to signal strength and geometric decay in representations of wireless networks.
Some authors have sought to specialize SINR-like settings to the Euclidean space, in order to leverage its specific topological properties, as for example,  Avin et al.~\cite{AvinEKLPR12} in their work on the SINR diagrams.
Others have sought to abstract from the coordinates in a plane but maintain a metric space of bounded growth that defines distances, like in the papers of Daum et al.~\cite{DGKN13} and Jurdzi\'nski et al.~\cite{JurdzinskiKRS-PODC14}.
Still others, like Bodlaender and Halld{\'o}rsson~\cite{BodlaenderH13} and Halld{\'o}rsson et al.~\cite{HalldorssonTWY16}, attempted to abstract the geometry even further and consider models that relax the properties of distance as represented by metric spaces.

\section{Technical Preliminaries}

\label{sec:technical-preliminaries}

Executions of algorithms are synchronous, in that they are structured as sequences of rounds of equal duration.
The size of messages transmitted by nodes is scaled to the length of a round, so that one message can be transmitted and received in a round.
We do not assume the existence of a global clock giving the same consecutive round numbers to each node.

There are $n$ nodes in a network.
Nodes have unique names assigned to them.
Each node's name is an integer in the range $[1,N]$, for some positive~$N$.
We treat the numbers $n$ and~$N$ as independent parameters in formulas, subject only to the assumption that $N\ge n$.

A broadcast algorithm disseminates some contents throughout the network; this contents is referred to as a \emph{rumor}.
Rumors and messages are also mutually scaled to each other such that a message  consists of at most one rumor.
A message may contain a sequence of control bits, whenever needed, to help nodes coordinate their concurrent actions.
This set of restrictions on communication is often referred to in the literature as of \emph{separate} or \emph{bounded} messages, which comes from the interpretation that multiple rumors require multiple  messages, a separate message per each rumor. 
The number of control bits per message in our algorithms is always~$\cO(\log N)$.

\Paragraph{Models of wireless communication.}

Networks are embedded in a two-dimensional Euclidean space.
Each node can be identified by its coordinates in a Cartesian coordinate system.
These  coordinates determine the Euclidean distance $\dist(x,y)$ between any pair of  points $x$ and~$y$.

The  SINR interference model involves the following parameters:
 \emph{path loss $\alpha> 2$},  \emph{ambient noise~$\cN > 0$},  \emph{transmission success threshold $\beta >0$}, \emph{sensitivity $\varepsilon_s$} and  \emph{connectivity  $\varepsilon_c$} such that $0\le \varepsilon_s\le \varepsilon_c<1$.
The \emph{transmission strength} of a node~$v$  is a positive real number denoted by~$P_v$.

If node $v$ transmits and $u$ is a different node then the \emph{signal strength} of $v$'s transmission as reaching~$u$ is denoted by~$\cP(v,u)$ and is defined as follows:
\begin{equation}
\label{eqn:signal-strength}
\cP(v,u) = P_v\cdot\dist(v,u)^{-\alpha}
\ .
\end{equation}
We will consider only \emph{uniform} networks in which all transmission strengths of nodes are equal, and denoted as~$P$.

Let $\cT$ be a set of nodes that transmit together in a round.
For any two nodes $v$ and $u$, where $v \in \cT$ and $u \notin \cT$, the \emph{interference at~$u$} of this configuration is denoted by $\cI(v,u,\cT)$ and is defined as follows:
\begin{equation}
\label{eqn:interference-strength}
\cI(v,u,\cT)=\sum_{w\in\cT\setminus\{v\}} \cP(w,u) 
\ .
\end{equation}
Observe that $\cI(v,u,\cT)=0$ if and only if $\cT=\{v\}$.
The \emph{signal-to-interference-plus-noise ratio} in this configuration is denoted by SINR\,$(v,u,\cT)$ and is defined as follows:
\begin{equation}
\label{eqn:sinr-definition}
\textrm{SINR}(v,u,\cT)
=
\frac{\cP(v,u) }{\cN+\cI(v,u,\cT)}
\ .
\end{equation}

We say that \emph{node~$u$ hears node~$v$} in a round when the following holds: (1) $v$ transmits in this round, (2) $u$ does not transmit in  this round, and (3) $u$ successfully  receives the message transmitted by~$v$.
A node $u$ is in the \emph{hearing range of node~$v$} if $u$ can hear a message transmitted by $v$ in a round in which $v$ is the only node in the network that transmits.
Nodes execute algorithms driven by the following two kinds of events only: either hearing a whole message from a node or not hearing a whole message from a node.
Nodes do not react to any other medium-sensing feedback from the wireless network.

Next, we explain the categorization of models with respect of sensitivity, which determines when nodes can hear transmitted messages.
Again, let a set of nodes~$\cT$ consist of these nodes that transmit in a round, and let nodes $v$ and~$u$ be such that $v \in \cT$ and $u \notin \cT$.

In the model of \emph{strong sensitivity}, a node~$u$ hears node~$v$ in this round when the inequality $\textrm{SINR}(v,u,\cT) \ge \beta$ holds, where parameter $\beta >0$ is a transmission success threshold.
This condition determines the hearing range of~$v$ as a distance from~$v$, which can be determined as follows.
A distance $d$ from which a message transmitted by a node~$v$ can be heard by~$u$  is determined by the formula~\eqref{eqn:sinr-definition} and satisfies the inequality 
\begin{equation}
\label{eqn:hearing}
\frac{P \cdot d^{-\alpha} }{\cN+\cI(v,u,\cT)} \ge \beta
\ .
\end{equation}
The maximum magnitude of such a distance~$d$ can be found as follows.
The ambient noise~$\cN$ is a fixed part of the left-hand side of~\eqref{eqn:hearing}, but the interference  part~\eqref{eqn:interference-strength} can vary and it is minimized when $\cI(v,u,\cT)=0$, which holds when $\cT=\{v\}$.
It follows that the maximum $d$ attainable in~\eqref{eqn:hearing} is determined by setting  $\cI(v,u,\cT)$ to~$0$  and equals $ (P/(\cN\beta))^{1/\alpha}$; we call this quantity the \emph{network radius} and denote by~$r$.
For the strong-sensitivity model, the meaning of the network radius~$r$ is such that a node~$u$ is in the hearing range of $v$ if and only if $d(v,u)\le r$.
In uniform networks, the network radius is a number determined by a network that does not depend on a transmitting node, while acting as a single transmitter, nor on listening nodes.

We may remark that if $\textrm{SINR}(v,u,\cT) \ge \beta$  then also 
\[
\cP(v,u) >  \beta \cdot \cI(v,u,\cT) = \beta\cdot \sum_{w\in\cT\setminus\{v\}} \cP(w,u)
\ ,
\]
because $\cN>0$. 
It follows that, in the model of strong sensitivity,  if $\beta\ge 1$ and a node $u\notin\cT$ hears a node~$v$ then node~$u$ cannot hear any other node in~$\cT$.
In this paper we assume only that $\beta>0$.

In the model of \emph{weak sensitivity}, a node~$u$ hears another node~$v$ in a round when $v$ transmits and  $\textrm{SINR}(v,u,\cT) \ge \beta$ and $\dist(v, u) \le (1 - \varepsilon_s)\cdot r$, where $r=(P/(\cN\beta))^{1/\alpha}$ is the network radius again.
We use the notation $R=(1 - \varepsilon_s)\cdot r$ and call $R$ the \emph{hearing radius}.

The  categorization of models into two classes of weak and strong sensitivities (devices) was introduced by Jurdzi\'nski et al.~\cite{JKS-ICALP-13}.
Weak sensitivity may be justified by the fact that it is often too costly for wireless devices to perform signal acquisition continuously, due to the constraints of the technology of wireless communication,  see Goldsmith and Wicker~\cite{GW-WN-02}. 
An alternative is  to wait for an energy spike, as represented by the condition SINR\,$(v,u,\cT) \ge \beta$.
Once nodes experience it, they may start sampling and correlating to synchronize and acquire a potential packet preamble, see Schmid and Wattenhofer~\cite{SchmidW06}. 
After that they can detect signals as determined by the formulas of weak sensitivity.
Observe that  if a node wants to transmit to a distance of the network's radius~$r$, then this node needs to be the only transmitting one in a network, but when a range of desired transmission is restricted to $(1 - \varepsilon_s)\cdot r $, then several nodes may succeed in transmitting concurrently.
In this paper, we use the model of weak sensitivity.

There is a natural algorithmic interpretation of weak sensitivity for wireless networks operating under the SINR regime.
Namely, such a network could be understood as a unit disc graph with transmission range~$R$, where $R$ is the hearing radius, which  is further restricted by the property  that only a ``dominating'' signal from a station within the transmission range can be heard. 
One may observe that algorithms for weak-sensitivity wireless networks could employ techniques similar to the ones used in unit disc  graphs to develop transmission schemes relying on sets of transmitters positioned relatively ``sparsely'' throughout the network. 
Guided by this observation, we use strongly selective families with suitably chosen parameters to implement message exchanges among the nodes.

\Paragraph{Graphs.}

Let $G=(V,E)$ be a simple graph.
A set of vertices $M\subseteq V$  is \emph{independent} when no two vertices in $M$ are connected by an edge in~$E$.
An independent set of vertices of $G$ is \emph{maximal independent} (or is a \emph{MIS}) when it is maximal in the sense of inclusion among all independent sets.
A set of vertices $B\subseteq V$ is \emph{dominating} in $G$ when every vertex not in $B$ is adjacent to at least one vertex of~$B$.
A dominating set in \emph{minimal dominating} when it is minimal in the sense of inclusion among all dominating sets.
A maximal independent set in~$G$ is also a minimal dominating set of~$G$.

We consider the length of a path to be the number of hops, so the length of each edge is~$1$.
A shortest path connecting two members of a maximal independent set that does not include at least one other member of the maximal independent set has length at most three, and the number three is smallest  as such a bound in general.
In other words, if there are at least two members in an independent  set, then a vertex in this independent set is connected to some other member by a path of length at most three.

\Paragraph{Communication graphs.}

A wireless network differs from wireline ones by lacking physical links.
Wires used as links allow to interpret the network as a graph with nodes acting as vertices and links serving as edges.
Associating a graph with a network helps to interpret communication algorithms as working in a clean abstract model of graphs, where information flows through edges.
A similar benefit could be obtained for wireless networks by associating graphs with them.
Such graphs are called ``communication graphs'' in this paper.
The nodes serve the purpose to be vertices of communication graphs, but what determines edges is less apparent.

The \emph{communication graph} of a wireless network is defined as follows: all the network's nodes are its vertices, and  for any two nodes $u$ and $v$, they are connected by an edge in the communication graph when the inequality $\dist(u,v)\le (1-\varepsilon_c) \cdot r$ holds, where $r$ is the network's radius and a connectivity coefficient $\varepsilon_c$ satisfies $0\le \varepsilon_s\le \varepsilon_c<1$.
We say that nodes $u$ and $v$ are \emph{$k$-hops away} from each other, or are \emph{$k$-hop neighbors}, when $k$ is the length of a shortest path connecting $u$ to $v$ in the communication graph.
A simultaneous transmission of two or more neighbors of a node~$v$ is called a {\em collision} at this node~$v$.
A collision at $v$ does not produce any special medium-sensing feedback at~$v$, but results in no message successfully received, for $\beta\ge 1/2+1/\cN$.

If $\varepsilon_c=\varepsilon_s$ then the model is of \emph{weak connectivity}, and if $\varepsilon_c>\varepsilon_s$ then the model is of  \emph{strong connectivity}.
In the model of weak connectivity, any two nodes $u$ and $v$ are connected by an edge in the communication graph if the inequality $\dist(u,v)\le R$ holds, where $R$ is the hearing radius.

The categorization of wireless networks with respect to weak and strong connectivity was introduced by Daum et al.~\cite{DGKN13}.
The problem of broadcasting allows to differentiate between the two models.
Strong connectivity allows to develop a broadcast algorithm of running time that depends linearly on the diameter~$D$ while other parameters contribute sublinear factors, see Table~\ref{tab:1}.
In contrast to that, weak connectivity demand time that is either $\Omega(n)$ or $\Omega(\min\{D\Delta,n\})$, depending on sensitivity.

We assume  the models of weak sensitivity and weak connectivity and use hearing and communication graphs as determined by these models.
To simplify notations, we denote the parameter $\varepsilon_c=\varepsilon_s$ by~$\varepsilon$.

\Paragraph{Grids and boxes.}

\begin{figure}[t]
\begin{center}
\includegraphics[width=0.40\textwidth]{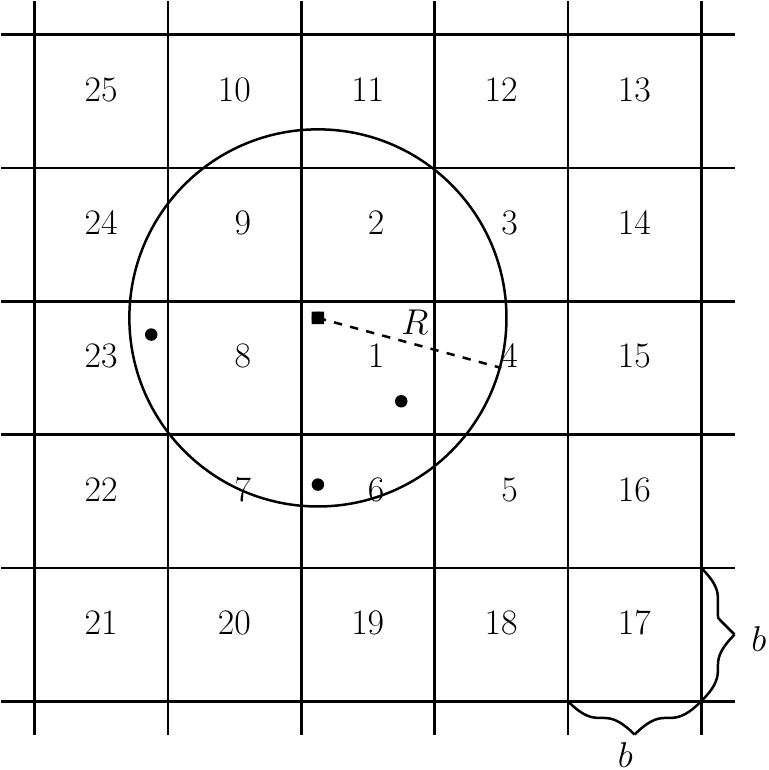}

\caption{\label{fig:grid} 
A depiction of a pivotal grid.
The circle of radius $R$ centered at the square dot point is the hearing range of this point.
The round dot point in box~$1$ belongs to the same box as the center of the circle.
The dot point in box~$6$ belongs to the $3\times 3$ square centered at the box containing the center of the circle.
The dot point in box~$23$ belongs to the $5\times 5$ square centered at the box containing the center of the circle.}
\end{center}
\end{figure}

Communication algorithms for wireless networks with nodes interpreted as points in a Euclidean space may be designed to leverage the geometric properties of this space.
For a constant $b>0$, we consider a grid of lines parallel to the coordinate axes that partition the space into $b\times b$ disjoint boxes, and such that $(0,0)$ is an intersection of a horizontal line and a vertical one; see Figure~\ref{fig:grid} for an illustration.
Specifically, a \emph{box} determined by its internal point includes the points on its left side without the top endpoint and the points on its bottom side without the rightmost endpoint, and the other points on the boundary are excluded.
Two boxes are \emph{adjacent} when their interiors are separated by a line of the grid.
Two boxes \emph{share a corner} when they share exactly one point on their boundaries.
The \emph{grid distance} between two boxes is understood in the Manhattan-metric sense, in that it is a natural number equal to the minimum number of hops between two adjacent boxes needed to move from one of them to the other.
In particular, adjacent boxes are of distance~$1$, and there are $4j$ boxes of grid distance $j$ from any box.


If $b=R/\sqrt{2}$, where $R=(1-\varepsilon)\cdot r$ is the hearing radius, then the grid is called \emph{pivotal}.
We will use only the pivotal grid in this paper.
Any two nodes in a box of the pivotal grid are within the Euclidean distance~$R$ from each other.
It follows that there are at most $\Delta+1$ nodes in a box, since all these nodes induce a clique and $\Delta$ is an upper bound on the degree of a vertex in a communication graph.

All the neighbors of a vertex~$v$ belong to $25$ boxes in the ``$5\times 5$ square'' of boxes centered at the box of~$v$.
This is depicted in Figure~\ref{fig:grid}, where the square dot represents such a vertex~$v$.
To verify this, observe that one hop in horizontal direction from a vertical edge of a box covers distance at most~$R$, which is less than the width of two columns of boxes, because $R<\sqrt{2} R=2\cdot \frac{R}{\sqrt{2}}$, while $5=2+1+2$.

All the vertices $2$-hop away from a vertex~$v$ belong to $49$ boxes in the ``$7\times 7$ square'' of boxes centered at the box containing~$v$.
To verify this, observe that two hops in horizontal direction from a vertical edge of a box covers distance at most $2R$, which is less than the width of three columns of boxes, because $2R < 3\cdot \frac{R}{\sqrt{2}}$, and we have $7=3+1+3$.

All the vertices $3$-hop away from a vertex~$v$ belong to $121$ boxes in the ``$11\times 11$ square'' of boxes centered at the box containing~$v$.
To see this, observe that three hops in horizontal direction from a vertical edge of a box cover distance at most $3R$, which is less than five columns of boxes, because $ 3R<5\cdot \frac{R}{\sqrt{2}}$, and $11=5+1+5$.

\Paragraph{Backbones.}

A backbone of a network is a subnetwork that facilitates global communication tasks, similarly as a spanning tree does.
Given a communication graph~$G$, its subgraph~$H$ that is a backbone of~$G$ is required to be connected, similarly to a tree, but rather than  spanning~$G$ it is a dominating set of~$G$.
A backbone is required to have asymptotically the same diameter as that of~$G$, so that implementing broadcast of a rumor by flooding inside a backbone does not incur an extra distance to cover.
At the same time, degrees of nodes in a backbone are required to be small, which facilitates collision resolution is wireless communication.
The notion is more involved though, because each node is additionally equipped with specialized algorithms that facilitate using a backbone efficiently. 
A precise specification is as follows.

Consider a network with a communication graph~$G$ of diameter~$D$.
A \emph{subnetwork~$H$ of $G$} is an induced subgraph of~$G$.
Backbones are subnetworks that have the suitable topological properties along with local algorithms associated with nodes.
We say that a subnetwork~$H$ is a \emph{backbone of~$G$} when it has the following topological properties:
\begin{enumerate}
\item
The nodes of $H$ form a connected dominating set of $G$.
\item
Each node's degree in $H$ is $\cO(1)$.
\item
The diameter of $H$ is $\cO(D)$.
\item
For each node $v$ in $G\setminus H$, there is exactly one neighbor $w$ in~$H$  assigned to~$v$; the node $w$ is called a \emph{representative of~$v$} and $v$ is said to be \emph{associated with~$v$}.
\end{enumerate}
The definition above  is to mean a method of construction of backbones that has implicit constants in the asymptotic notation that do not depend on the size of the communication graph~$G$.
Such a method is also expected to assign representatives to neighbors of the obtained backbone in some systematic manner.

There are two \emph{local algorithms} \textsc{Intra$_H$} and \textsc{Inter$_H$} that provide communication functionality of a backbone~$H$ for a network~$G$, that implement intra-backbone communication among the nodes in the backbone and inter-backbone communication between nodes not in the backbone and their representatives in the backbone, respectively.
These  algorithms have the following functionality:
\begin{quote}
\begin{description}
\item[\rm Algorithm \textsc{Intra$_H$}:] 
It facilitates exchanging messages between each pair of neighbors in~$H$. 
In the process of executing \textsc{Inter$_H$}, each message received by a node $v\in H$ is also delivered to all the nodes in~$G$ associated with $v$, that is, to the nodes for which $v$ is a representative.
\FF
\item[\rm Algorithm \textsc{Inter$_H$}:]  
It  facilitates delivering a message from a node to its representative.
\end{description}
\end{quote}
The notion of a backbone we use is defined similarly as in Jurdzi{\'n}ski and Kowalski~\cite{JK-DISC12}. 
Backbones can be used as a generic tool for many communication and computation tasks in the network, see Halld{\'{o}}rsson and Tonoyan~\cite{HalldorssonT21}, Jurdzi{\'n}ski and Kowalski~\cite{JK-DISC12},  and Yu et al.~\cite{YuWWY2013}.

\Paragraph{Bare-bones algorithms.}

The following restrictions on algorithms for wireless communication make them \emph{bare-bones}: each node knows only its name and the numbers $N$ and $\Delta$, and additionally the size of messages is constrained such that a single message carries $\cO(\log N)$ bits.
The motivation for studying communication algorithms under bare-bones constraints is that the restrictions imposed on the algorithms make them easily portable and the obtained performance bounds widely applicable.

Algorithms with similar bare-bones restrictions were considered by Jurdzi{\'n}ski et al.~\cite{JurdzinskiKRS-PODC14}  in the case of strong-sensitivity and strong-connectivity of wireless networks, while we consider bare-bones algorithms for a combination of weak sensitivity and weak connectivity of wireless networks.
The algorithms in Jurdzi\'nski and Kowalski~\cite{JK-DISC12} and Jurdzi\'nski et al.~\cite{JKS-ICALP-13}, being also for weak-connectivity and weak-sensitivity of the wireless communication, were designed with the assumption that nodes know their coordinates in a plane; in this work we develop efficient algorithms that do not use this information.

\Paragraph{Rooted spanning trees.}

Consider a simple connected graph $G=(V,E)$ with a distinguished \emph{source} vertex.
The source generates a token which traverses the graph and in the process builds a subgraph.
Here ``traversal'' means that the token hops from a visited vertex to its neighbor by traversing the connecting edge.
The specific manner of token traversal and constructing the subgraph that we use is called \emph{breadth-then-depth} and is described as follows.

The source generates a token, and so becomes a first vertex visited by the token.
Vertices are categorized into \emph{discovered} and \emph{hidden}.
Initially, the source is discovered and every other vertex is hidden.
Suppose that the token visits a vertex~$v$.
If this is a first visit of the token then all the neighbors of $v$ that are still hidden become discovered and the edges connecting them to~$v$ become added to the constructed subgraph; the newly added vertices become the \emph{children} of~$v$ and vertex~$v$ becomes their \emph{parent}.
The source vertex is considered to be its own parent and is a \emph{root}.
The vertex $v$ dispatches the token to one of its children that has not been visited by the token yet, if such vertices exist.
If the token comes back again to~$v$ and there are still children of $v$ not visited by the token then $v$ dispatches the token to one of them, and otherwise sends the token back to its parent.
The traversal terminates when the source is about to send the token to itself. 
A tree produced during a breadth-then-depth traversal of a graph is called a \emph{breadth-then-depth} spanning tree.
Such a tree is rooted at the source.

Chlebus et al.~\cite{ChlebusKPR-ICALP11} used breadth-then-depth trees in their distributed algorithms in radio networks.
Next, we summarize the relevant propertied of breadth-then-depth traversal as Fact~\ref{fact:breadth-then-depth}.  


\begin{fact}
\label{fact:breadth-then-depth}

The breadth-then-depth traversal of a simple connected graph creates a spanning tree of the graph.
The token visits the vertices by traversing the obtained spanning tree  in a depth-first manner.
\end{fact}

\begin{proof}
The vertices and edges that are added to the constructed graph belong to the original traversed  graph, so the constructed graph is a subgraph.
We show next that the edges connecting children to parents make a subgraph that is connected, has no cycles, and it includes all the vertices.

Connectivity follows from the fact that edges traversed by a token  determine a walk, which creates a path to the source vertex after being pruned of repetitions of edges.
A cycle cannot occur because an edge connects two vertices such that when one of them was first visited by the token then it was already discovered while the other became its child because it was still hidden.
Since the created subgraph is connected and acyclic, it is a tree.

We show now that all vertices become discovered, and then they get connected as children to their respective parents. 
If there is only one vertex then it is the source and it belongs to the tree; otherwise let~$w$ be a vertex different from the source.
Let $v_1, \ldots,v_k$ be the children of the source, in the order in which the source sends the token to them. 
Observe that if a token is sent by the source to its child then it comes back to the source through the same child, as otherwise the token would traverse a cycle.
Let $v_i$ be such that $i$ is smallest with the property that there is a path connecting~$v_i$ with~$w$ that does not pass through the source: this path traverses only undiscovered vertices before the token visits~$v_i$.
When the token visits $v_i$ and starts traversing the graph then it will discover and visit $w$,  because the token will discover and visit all the vertices on the path to~$w$.
Since all the vertices are incorporated into the created tree, it is a spanning tree.

The traversal of the spanning tree, as it unfolds for the traveling token, is specified by the depth-first search traversal principle: the token explores all the discoverable vertices before it returns to the parent.
\end{proof}

\Paragraph{Sets and sequences of sets.}

For a natural number $N$, the notation $[N]$ denotes the set $\{1,2,\ldots,N\}$.
We consider subsets of $[N]$ and sequences of such subsets; the notation $\cS$ will denote a sequence of subsets of $[N]$. 
Such a sequence $\cS=(S_0,\ldots,S_{t-1})$ of subsets of~$[N]$ consisting of $t$ terms is said to be of \emph{length~$t$}. 
By writing $B\in \cS$ we mean that $B$ is a term of such a sequence $\cS$, as if $\cS$ were an unordered family of sets.

We identify a sequence of sets $\cS=(S_0,\ldots,S_{t-1})$ with a broadcast schedule~$\cS'$ consisting of $t$ consecutive rounds, in which a node $v$ transmits in round $i$ if and only if  $v\in S_i$, for $0\le i<t$.
Such a broadcast schedule is said to be performed by \emph{executing~$\cS$}.

Let $\cS$ be a sequence of subsets of $[N]$. 
For a subset $A\subseteq N$ and $a\in A$, we say that \emph{$a$ is selected from~$A$ by~$\cS$} if there is a set $S\in \cS$ such that $A\cap S=\{a\}$.

Let $N$, $x$ and $y$ be positive integers such that $y \leq x \leq N$. 
We say that a sequence~$\cS$ of subsets of~$[N]$ is a \emph{$(N, x, y)$-selector} if for each set $A \subseteq [N]$ of $x$ elements  there are at least $y$ elements in $A$ that can be selected from $A$ by $\cS$. 
For any fixed $\zeta$ such that $0<\zeta<1$, there is an $(N, x, \zeta x)$-selector of   size $\cO(x\log N)$, see~Chlebus and Kowalski~\cite{ChlebusK05} and DeBonis et al.~\cite{DeBonisGV05}.

For an integer $c>0$, a sequence~$\cS$ is \emph{$(N,c)$-strongly-selective} if, for every non-empty subset $Z$ of $[N]$ such that $|Z|\leq c$, and for each element $z\in Z$, this $z$ can be selected from $Z$ by~$\cS$. 
For each $c>0$ such that $c\leq N$, there exists an $(N,c)$-strongly-selective sequence of length $\cO(c^2\log N)$; 
see Clementi et al.~\cite{ClementiMS01}.

Whenever we use combinatorial structures such as selectors or strongly-selective sequences in algorithm design, then it is assumed that they are a part of code.

\Paragraph{Odds and ends.}

Performance bounds of an algorithm hold \emph{with high probability} if, for any constant $d \geq 1$, the performance bounds can be made to hold with  probability at least $1 - {n^{-d}}$ by suitably adjusting constants in a code of the algorithm.
We assume that the numbers $n$, $N$ and $\Delta$ are all powers of $2$, for the simplicity of exposition; in a general case, these parameters can be rounded up to the nearest power of~$2$.
 The notation $\lg x$ means logarithm of $x$ to the base~$2$.

\section{Algorithmic Tools}

\label{sec:algorithmic-tools}

We present building blocks of algorithms and concepts and tools used in deriving their performance bounds.

We propose a construction of an induced subgraph~$H$ of a simple graph~$G$ such that $H$ is connected dominating, which we call \emph{shortcut connecting}.
Start with a maximal independent set~$M$ as an initial dominating set; it will grow to eventually  make the  vertices of a connected subgraph~$H$.
We grow $H$ by iterating the following process.
Suppose that there exist two vertices $v_1$ and $v_2$ in $M$ that are connected in $G$ by a path of length at most three but they are not connected in $H$ by a path of length at most three: add to $H$ either one or two vertices along with all the newly induced edges that provide a missing shortcut such that $v_1$ and $v_2$ are now connected in $H$ by a path of length at most three.
The vertices added this way are called \emph{connectors} for $M$.
The final graph $H$ consists of the initial maximal independent set $M$ and all the added connectors.


\begin{proposition}
\label{pro:shortcut-connecting-general}

For a simple connected graph $G=(V,E)$ of diameter $D$ and a maximal independent set of vertices $M\subseteq V$, if an induced subgraph $H$ is obtained from $M$ by shortcut connecting then $H$ is connected,  its vertices make a dominating set of $G$, and the diameter of $H$ is at most $3D+2$.
\end{proposition}

\begin{proof}
The set of vertices of the subgraph $H$ is dominating in~$G$ because $M$ is already such.
To show it is connected, suppose otherwise, to arrive at a contradiction.
Each connected component of $H$ includes at least one element of~$M$.
This is because otherwise adding any single vertex from this component to $M$ would make it larger and still independent.
Consider a shortest path in $G$ connecting two vertices $v_1$ and $v_2$ in $M$ that are in different connected components of~$H$.
This shortest path does not include vertices from~$M$, by its minimality. 
So $v_1$ and $v_2$ are connected by a path in~$H$ of length at most three.
This means that $H$ is not the final subgraph produced by shortcut connecting, which is a contradiction.

Next we estimate the diameter of~$H$.
Take a simple path $P=(v_1,\ldots,v_k)$ in~$G$ that has both endpoints in~$H$.
For each vertex $v_i$ on $P$ if $v_i\in M$ then denote $v_i$ also by $w_i$ and otherwise if $v_i$ does not belong to $M$ then let $w_i$ be a neighbor of $v_i$ that belongs to~$M$.
Consider a sequence of vertices $(w_1,\ldots,w_k)$.
For each pair $w_i, w_{i+1}$, these vertices are connected by a path of length at most three in $G$, by their selection.
Since both $w_i$ and $w_{i+1}$ are in $M$, they are connected by a path of length at most three in $H$.
It follows that for each path in $G$ connecting two vertices in $H$ of length $L$ there exists a path in $H$ connecting the same pair of vertices of length at most $3L+2$.
\end{proof}

The following Proposition~\ref{prop:signal-strength} gives a useful estimate of signal strength with our assumptions about the model of communication.


\begin{proposition}
\label{prop:signal-strength}

In the model of weak sensitivity and weak connectivity, if a node $v$ transmits and another node~$u$ is such that $u$ and $v$ are neighbors in the communication graph then the signal strength at $u$ of this transmission is at least
$(1+\varepsilon)\cN\beta$.
\end{proposition}

\begin{proof}
The signal strength $\cP(v,u)$ at~$u$ of a transmission is given by the formula~\eqref{eqn:signal-strength}.
The distance between $u$ and $v$ is at most $R$, by the specification of edges of the communication graph.
The signal strength can be estimated as follows:
\begin{equation}
\label{eqn:signal}
\cP(v,u)\ge P \cdot R^{-\alpha} 
= P\cdot  (1-\varepsilon)^{-\alpha}(P/(\cN\beta))^{1/\alpha})^{-\alpha} \ge (1+\varepsilon)\cN\beta
\ ,
\end{equation}
where we used $(1-\varepsilon)^{-\alpha}>1+\varepsilon$ for $0<\varepsilon <1$ and $\alpha\ge 1$.
\end{proof}

The communication graph could be discovered by the nodes of a wireless network by having them transmit one by one in a systematic manner: when a node~$u$ hears a message from another node~$v$ that is the only transmitter in the whole network then it may come from a neighbor in the communication graph.
To determine if the inequality $\dist(u,v)\le (1-\varepsilon_c) \cdot r$ holds, the node $u$ could compute the distance $\dist(u,v)$ by resorting to the nodes'  coordinates, assuming the nodes know their own coordinates and include them in transmitted messages as a ``signature'' identifying the sender, and also that they know the network radius~$r=(P/(\cN\beta))^{1/\alpha}$ along with the connectivity coefficient~$\varepsilon_c$.
After a preprocessing that allows all the nodes to learn and remember their neighbors in the communication graph, this could be used in distributed algorithms to construct subgraphs of the communication graph (like backbone) when such algorithms rely on neighbors in the communication graph exchanging messages.
Namely, when a message is heard and it does not come from a neighbor in the communication graph then it could be ignored.
A simpler approach suffices in the model of weak sensitivity and weak connectivity, as is summarized in the following Proposition~\ref{prop:hearing-only-from-neighbors}.
This fact justifies why nodes executing algorithms we develop need to know so little, and in particular, they do not need to know their positions in a system of coordinates.


\begin{proposition}
\label{prop:hearing-only-from-neighbors}

In the model of weak sensitivity and weak connectivity, if a node $u$ hears a message transmitted by a node $v$ in a round then $u$ and $v$ are neighbors in the communication graph.
\end{proposition}

\begin{proof}
Let $\cT$ be the set of nodes transmitting in a round, where $v\in\cT$.
The node~$u$ hears the message transmitted by~$v$ if both inequalities 
$\textrm{SINR}(v,u,\cT) \ge \beta$ and $\dist(v, u) \le R$ hold, by weak sensitivity.
The nodes $u$ and $v$ are neighbors in the communication graph when the inequality $\dist(v, u) \le R$ holds, by weak connectivity.
\end{proof}

\Paragraph{Shortcut connecting in communication graphs.}

We obtain a connected dominating set by adding certain vertices to an independent set of a wireless network.
The resulting subgraph of the communication graph has properties useful for implementing communication algorithms in wireless settings.


\begin{proposition}
\label{pro:shortcut-connecting-communication-graphs}

For a connected communication graph $G=(V,E)$ of a wireless network and a maximal independent set of vertices $M\subseteq V$ in~$G$, if a subgraph~$H$ of~$G$ is obtained from~$M$ by shortcut connecting then there is a constant upper bound on node degrees of~$H$ and the number of nodes in~$H$ is $\cO(m)$, where $m$ is the size of the smallest connected  subgraph of~$G$ dominating in~$G$.
\end{proposition}

\begin{proof}
Each box of the pivotal grid contains at most one member of~$M$.
A node has neighbors in at most $25$ boxes, including its own.
A member of $M$ is connected via connectors to at most $24$ other members of~$M$.
A node that is a connector plays this role for at most $24\times 25$ pairs of members of~$M$.
It follows that there is a constant upper bound on the number of nodes in $H$ in one box of the pivotal grid.
This gives a constant upper bound on degrees of every node of~$H$.

Let $K$ be an arbitrary set of  nodes of~$G$ such that $K$ is dominating, the subgraph induced by~$K$ in~$G$ is connected, and the size  of~$K$ is smallest with these two properties.
We observe that each node $v$ of~$K$ is connected to a constant number of nodes in~$H$. This is because the neighbors of $v$ in $H$ have to belong to at most $25$ boxes  of the pivotal grid, determined by the box of~$v$, and there is a constant upper bound on the number of nodes in~$H$ in one box of the pivotal grid.
It follows that the size of $H$ is at most a constant multiple of the size of~$K$.
\end{proof}

Next we discuss the efficiency of using strongly selective sequences to facilitate communication among neighbors in a communication graph.
The following Proposition~\ref{prop:SINR-and-boxes} states a critical technical insight that we will use to reason about properties of wireless communication in Euclidean space with the weak sensitivity and weak connectivity.
A similar approach was applied in Jurdzi\'nski et al.~\cite{JKS-ICALP-13}. 
We assume that some nodes are \emph{active} in that only they can transmit.


\begin{proposition}
\label{prop:SINR-and-boxes}

In the model of weak sensitivity and weak connectivity, if there are at most $z$ active nodes per box, then there is a number~$c$ such that $c=\Theta(z^3)$ and with the property that if all the active nodes execute an $(N,c)$-strongly-selective sequence, then, for each active node~$v$, there is a round in which the node~$v$  is heard by all its neighbors in the communication graph.
\end{proposition}

\begin{proof}
Let us define $\eta=\delta z/(\varepsilon\cN)$, where $z$ is smallest such that there are at most $z$ active nodes in each box and $\delta>0$ is a constant parameter to be determined later.
The quantity~$\eta$ is a linear function of $z$, since $\varepsilon$ and $\cN$ are both fixed parameters in wireless networks.

Consider an active node~$v$ in a network under consideration.
Let the other active nodes in the boxes within the grid distance of at most $\eta$  of the box of node~$v$ make a set~$A$.
The number of such active nodes is $|A|=\cO(z \eta^2)$. 
Let $\delta$ be large enough so that all the neighbors of $v$ are in~$A$.

We want a strongly selective sequence such that when it is executed then there is a round in which $v$ transmits but none of the nodes in $A$ does.
To this end, it is sufficient to take an $(N,c)$-strongly-selective sequence, where $c=K z\eta^2$ with a sufficiently large constant $K>0$.
Let $t$ denote this round, counting from the first round of executing the strongly-selective sequence.

We estimate the total interference at a neighbor of $v$ coming from the active nodes that are not in~$A$.
There are $4j$ boxes of distance $j>0$ from the box of~$v$, each with at most $z$ nodes transmitting among the active nodes.
The signal strength  from each of these nodes, at any node that is a neighbor of~$v$,  is $\Theta(j^{-\alpha})$.
The total interference at a neighbor of~$v$ from the transmitting nodes not in~$A$, as specified in~\eqref{eqn:interference-strength}, is bounded from above by a quantity proportional to 
\begin{equation}
\label{eqn:total-interference}
z \sum_{j> \eta} j\cdot j^{-\alpha}
=
\cO\bigl(z \eta^{1-\alpha } \cdot \sum_{j\ge 1} j^{1-\alpha}\bigr) 
=
\cO(z\eta^{1-\alpha}) 
\ ,
\end{equation}
by the estimate $\sum_{j\ge 1} j^{1-\alpha}=\cO(1)$, for $\alpha>2$.
The total interference at a neighbor of~$v$, as expressed by~\eqref{eqn:total-interference}, could be made  smaller than $\varepsilon\cN$ by choosing a sufficiently large~$\delta$ in the specification of~$\eta$, since $1/\eta=\varepsilon\cN/(\delta z)$.
It follows that the total interference plus noise at such a node is at most $\cN+\varepsilon\cN=(1+\varepsilon)\cN$.

The signal strength of a neighbor of $v$ in the communication graph is at least $(1+\varepsilon)\cN\beta$, by Proposition~\ref{prop:signal-strength}.
The value of SINR at such a node, according to the defining Equation~\eqref{eqn:sinr-definition}, is at least $\beta$.
This means that the conditions of hearing are satisfied, so that  node~$v$ is heard by all its neighbors in the communication graph in round~$t$.
\end{proof}

In applying Proposition~\ref{prop:SINR-and-boxes}, we will use $(N,c)$-strongly-selective sequence for sufficiently large~$c$, with $c$ being either $c=O(1)$ or $c=\Theta(\log^3 N)$, depending on whether $z=\cO(1)$ or $z=\cO(\log N)$, respectively, and of the respective lengths $\cO(\log N)$ and $\cO(\log^7 N)$.

\section{Broadcasting to Coordinate Start}

\label{sec:single-node-start}

We present a randomized algorithm for a single-source broadcast.
Nodes other than the source join an execution only after they get activated by receiving messages. 
A ``rumor'' that the source wants to disseminate among all the nodes fits into a message that can be transmitted in one round and consists of $\cO(\log N)$ bits of information.

The broadcast algorithm we develop may be used to synchronize a network.
This can be accomplished by forwarding a counter of rounds along with each message generated  by the broadcast algorithm.
Such a counter is inherited by each awoken node and is incremented with each round of broadcasting.
A predetermined threshold for the counter values may be established, determined by the running time of broadcast, such that when the counter reaches this threshold then this indicates reaching a synchronized start.
Once  such a synchronized-start round is reached, an algorithm designed for a synchronized start can be invoked simultaneously by all the nodes, like the backbone algorithm given in Section~\ref{sec:coordinated-start}.

The broadcast algorithm involves a token traversing the network by hopping along the edges of its communication graph. 
The token is initiated by the source node and performs a breadth-then-depth traversal, as summarized in Fact~\ref{fact:breadth-then-depth} in Section~\ref{sec:technical-preliminaries}.
Nodes do not know their neighbors in the communication graph when the traversal starts and they need to learn them to be able to control the moves of the token.
Discovering edges of the communication graph and sending a token across them is accomplished by sending messages, so the token traverses the edges of the communication graph of the wireless network, by Proposition~\ref{prop:hearing-only-from-neighbors}.

We will use two auxiliary randomized distributed routines to coordinate movements of the token.
They will be used to implement a breadth-then-depth traversal and we use the relevant terminology as explained in Section~\ref{sec:algorithmic-tools}.
One is to estimate the number of hidden neighbors of a node in the communication graph.
The other is to discover the hidden neighbors, based on knowing an estimate of their number.
The two auxiliary algorithms are discussed in the next two Subsections.

\subsection{Estimating the number of hidden neighbors}

\label{subsec:estimate-hidden-neighbors}

We discuss a randomized procedure to estimate the number of hidden neighbors.
The procedure is initiated and coordinated by a node, denoted~$s$, when it holds the token in the course of its  traversal of the network.
Only the nodes of distances at most $R$ from~$S$ participate, where $R$ is the hearing radius, since this distance determines neighborhoods in the communication graph in the weak-connectivity case. 
The procedure is called \textsc{Estimate-Hidden}.
Its pseudocode is in Figure~\ref{alg:estimate-hidden}.

The node~$s$ estimates the number of hidden neighbors by counting messages it hears and comparing the outcome to some threshold value.
The hidden neighbors of the node $s$ execute $1+\lg N$ stages. 
In stage~$i$, a hidden neighbor iterates the inner loop $d\lg N$ times.
A node transmits in one iteration of the inner loop in stage~$i$ with probability~$2^{-i}$, independently from other nodes.
If $k$ is the latest stage for which the number of messages heard by $s$ is at least $d \lg N \cdot 2^{-4}$ then node~$s$ considers the number~$2^{k}$ to be an upper bound  on the number of hidden neighbors.


\begin{figure}[t]

\hrule

\FF

\noindent
\texttt{procedure} \textsc{Estimate-Hidden} 

\FF

\hrule

\FF

\begin{enumerate}[nosep]
\item
$s$ initializes $k\gets 0$
\item
$s$ transmits a message inviting hidden neighbors to participate 
\item
\texttt{for} $i\gets 0$ \texttt{to} $\lg N$ \texttt{do} ~ ~ ~ /$\ast$ stage $i$ $\ast$/
\begin{itemize}[nosep]
\item[]
\texttt{if} $v$ is a hidden neighbor of $s$ \texttt{then}
\begin{itemize}[nosep]
\item[]
\texttt{for} $j\gets 1$ \texttt{to} $d \lg N$ \texttt{do} 
\begin{itemize}[nosep]
\item[]
$v$ carries out a random trial with probability $2^{-i}$ of success
\item[]
\texttt{if} a success occurs \texttt{then} $v$ \texttt{transmits}
\end{itemize}
\end{itemize}
\item[]
\texttt{if} $s$ hears a message at least $2^{-5} \cdot d \lg N$ times in this stage
\texttt{then} $k\gets i$
\end{itemize}
\item
\texttt{return} $2^{k}$ at $s$
\end{enumerate}

\FF

\hrule

\caption{\label{alg:estimate-hidden}
Pseudocode for a node $s$ and its neighbor $v$.
Transmissions of a dummy message are performed by $v$ and heard and counted by~$s$.
Constant $d$ is a parameter to be determined in analysis.
The number~$2^k$ returned by $s$ is interpreted as an approximation  of the number of hidden neighbors of $s$.}
\end{figure}


\begin{lemma}
\label{lem:hidden}

For each $a>0$, if $\rho>1$ is the number of hidden neighbors of $s$ then the number~$2^k$ returned by algorithm \textsc{Estimate-Hidden} satisfies  $\rho\le 2^k \le 2^5\cdot \rho$ with probability at least $1-n^{-a}$, for a sufficiently large parameter $d>0$ and all sufficiently large $n$.
\end{lemma}

\begin{proof}  
The procedure operates by nodes exchanging messages, which always arrive at a node from neighbors in the communication graph, by Proposition~\ref{prop:hearing-only-from-neighbors}.

Let the interval of integers $[2,2^{1+\lg N}-1]$ be partitioned into disjoint segments as follows:
\[
[2,3], [4,7], \dots, [2^{i},2^{i+1}-1], \dots, [2^{\lg N}, 2^{1+\lg N}-1]
\ .
\] 
The number $\rho$ is in precisely one of these ranges.
Let $\ell$ be such that $\rho \in [2^{\ell}, 2^{\ell+1}-1]$, where $\ell\ge 1$.

The probability that a specific hidden neighbor of $s$ transmits during stage $\ell$, while the other hidden neighbors do not transmit, can be estimated as follows: 
\[
\frac{1}{2^\ell}\Bigl(1-\frac{1}{2^\ell}\Bigr)^{\rho-1} 
\geq 
\frac{1}{2^\ell} \cdot \Bigl(1 - \frac{1}{2^\ell}\Bigr)^{2^\ell \frac{\rho}{2^\ell}}
\geq
\frac{1}{2^\ell} \cdot \Bigl(1 - \frac{1}{2^\ell}\Bigr)^{2^{\ell+1}}
\geq
\frac{1}{2^\ell} \cdot \frac{1}{16}
\ ,
\] 
because $\rho < 2^{\ell+1}$.
The probability that exactly one hidden neighbor of~$s$ transmits is  at least  
\begin{equation}
\label{eqn:probability-one-eight}
\rho \cdot \frac{1}{2^\ell} \cdot \frac{1}{16} \geq \frac{1}{2^4}
\ ,
\end{equation}
because  $\rho\ge 2^{\ell}$.
Let us define an indicator random variable $X_i^j$ such that $X_i^j = 1$ when a single transmission occurs in trial~$j$ of stage~$i$, otherwise $X_i^j = 0$. 
The estimate of~\eqref{eqn:probability-one-eight} can be interpreted to mean that  $\Pr (X_\ell^j=1) \geq \frac{1}{2^4}$.
Define a random variable $X_i$ as follows:
\[
X_i = \sum_{j \in [d \lg N]} X_i^j
\ . 
\]
By the linearity of expectation, the following inequality holds: 
\[
\mE[X_\ell] \geq 2^{-4} \cdot d \lg N
\ .
\] 
By the Chernoff bound, the number of successful transmissions received by node~$s$ in  stage~$\ell$ of algorithm \textsc{Estimate-Hidden} is smaller than $2^{-5} d \lg N$ with probability that is at most $n^{-2a}$, for sufficiently large~$d$.
It follows that the inequality $2^{k}\ge \rho$ does not hold with a probability that is at most $n^{-2a}$.

Consider a stage~$i$ of the algorithm such that $\ell+7\le i \le 1+\lg N$.
In a single round of this stage, the probability that there is at least one transmitting node is at most 
\[
\rho\cdot 2^{-i}
\le 
\rho\cdot 2^{-\ell-7}
<
2^{-6}
\ ,
\]
because $\rho<2^{\ell+1}$.
Thus the expected number of rounds with at least one transmission in stage~$i$ is smaller than  
$d \lg N \cdot 2^{-6}$.
By the Chernoff bound, the number of times  node~$s$ hears a message in stage~$i$ is at least $d \lg N \cdot 2^{-5}$ with a probability at most $n^{-3a}$, for a sufficiently large~$d$.
The number of such stages is at most $\lg N$.
By the union bound, some of these stages result in producing at least $d \lg N \cdot 2^{-5}$  messages that are heard with probability at most $\lg N\cdot n^{-3a}$, which is at most $n^{-2a}$ for sufficiently large~$n$.
Therefore, the inequality $2^{k} \le 2^5\rho$ does not hold  with probability at most $n^{-2a}$.

We conclude, by the union bound,  that both the inequalities $\rho\le 2^k \le 2^5\cdot \rho$   hold  with probability at least $1-n^{-a}$, for a sufficiently large parameter~$d$ and all sufficiently large~$n$.
\end{proof}


\begin{figure}[t]

\hrule

\FF

\texttt{procedure} \textsc{Discover$(x)$} 

\FF

\hrule

\FF

\begin{enumerate}[nosep]
\item
initialize list of discovered neighbors $R\leftarrow \emptyset$ at $s$
\item
\texttt{for} $j=0,1, \ldots, \lg x -1$ \texttt{do} ~ ~ ~ /$\ast$ stage $j$ $\ast$/
\begin{itemize}[nosep]
\item[]
notify neighbors $v$ of $s$: 
\begin{itemize}[nosep]
\item[]
\texttt{if} $v$ is hidden \texttt{then} $v$ starts executing $(N,2^{-j} x, 2^{-j-1} x)$-selector 
\end{itemize}
\item[]
during executing a selector by neighbors: 
\begin{itemize}[nosep]
\item[]
\texttt{if} a name $v$ is heard \texttt{then} $v$ gets discovered:  
node $s$ adds $v$ to $R$
\end{itemize}
\item[]
notify all neighbors discovered in this stage $j$: 
\begin{itemize}[nosep]
\item[] 
$s$ transmits their names one by one
\end{itemize}
\end{itemize}
\item
\texttt{return} $R$ as the list of discovered neighbors at $s$ 
\end{enumerate}

\FF

\hrule

\caption{\label{proc:Discover}
Pseudocode for a node~$s$.
The number $x$ is an upper bound on the number of hidden neighbors, it is a power of~$2$ and is passed as an argument.
The used selectors are of length $\cO(2^{-j} x\log N)$ and are part of code.
The discovered neighbors become children of~$s$ in a breadth-then-depth tree.}
\end{figure}

\subsection{Discovering hidden neighbors}

\label{subsec:discover-hidden-neighbors}

Let us assume that we have an upper bound $x$ on the size~$\rho$ of the hidden neighborhood of a node~$s$, where $\rho\le x\le 2^5 \rho$.
The node $s$ could obtain such an estimate~$x$ by executing procedure \textsc{Estimate-Hidden} presented in Section~\ref{subsec:estimate-hidden-neighbors}.
Given such a bound~$x$ for a node~$s$, a routine $\textsc{Discover}(x)$ allows the node to learn its neighborhood  in $\cO(x \log N)$ rounds, as we show next in Lemma~\ref{lem:Discover}. 

Pseudocode of procedure \textsc{Discover}  is given in Figure~\ref{proc:Discover}.
The procedure works by repeating \emph{stages} $\lg x$ times, each stage an iteration of the for-loop, with the goal to decrease  by half an estimate on the number of hidden neighbors.
This in turn is accomplished by having the hidden neighbors execute selectors of lengths determined by stage numbers.


\begin{lemma}
\label{lem:Discover}

If a node has at most $x$ hidden neighbors that an execution of $\textsc{Discover}(x)$ makes this node learn all its hidden neighbors in time $\cO(x\log N)$.
\end{lemma}

\begin{proof}
The procedure \textsc{Discover} operates by nodes exchanging messages.
A transmitted message always hops from a node to its neighbors in the communication graph, by Proposition~\ref{prop:hearing-only-from-neighbors}.

During executing procedure \textsc{Discover}$(x)$ by a node~$s$, at most $x/2$ nodes remain hidden after the first stage, byt the definition of selectors. 
This pattern continues, such that after a $j$th stage the number of hidden neighbors is at most~$x/2^{j-1}$.
Indeed, the proof is by induction on $j$. We just argued that it holds at the end of the execution of loop ``for'' applied
for $j=0$, thus assume that it holds for some $j\ge 0$. 
By the definition of $(N,2^{-j} x, 2^{-j-1} x)$-selector applied to the set of hidden neighbors, 
of size at most $x/2^{j}$ by induction, at most~$x/2^{j-1}$ of them remain unselected by the selector, hence hidden.
The lengths of these selectors decreases geometrically, so the lengths of the stages sum up to $\cO(x\log N)$.
\end{proof}

\subsection{Algorithm for broadcasting}
\label{sec:algorithm-traverse-to-broadcast}

We present an algorithm to have a token traverse the network.
The algorithm is called \textsc{Traverse-To-Broadcast}.
It is summarized as a pseudocode in Figure~\ref{alg:Traverse-To-Broadcast}.
The token can carry any contents piggybacked on it with the goal to broadcast it. 
A traversal is initialized by the node which is the source of a broadcast message. 
There is no prior coordination among the nodes of the network to participate in an execution.
The token's traversal involves building a breadth-then-depth spanning tree in the communication graph, rooted at the source node, and the token traverses it in a depth-first manner, according to Fact~\ref{fact:breadth-then-depth}.


\begin{figure}[t]

\hrule

\FF
 
\texttt{algorithm} \textsc{Traverse-To-Broadcast} 

\FF

\hrule

\FF

\begin{enumerate}[nosep]
\item
upon receiving and holding the token for the first time:  
\begin{itemize}[nosep]
\item[]
\texttt{if}  $s$ is different from the source \texttt{then} 
\begin{itemize}[nosep]
\item[]
record the name of the node from which the token arrived as  parent
\end{itemize}
\item[]
execute procedure \textsc{Estimate-Hidden} 
\begin{itemize}[nosep]
\item[]
\hfill 
{\em /$\ast$ returns an upper bound $x$ on the number of hidden neighbors $\ast$/}
\end{itemize}
\item[]
execute procedure \textsc{Discover}$(x)$ \hfill
{\em /$\ast$ returns a list $R$  of discovered neighbors $\ast$/}
\end{itemize}
\item
upon receiving and holding the token:
\begin{itemize}[nosep]
\item[]
\texttt{if} the token arrived from a node $v\in R$ \texttt{then} remove $v$ from $R$
\item[]
\texttt{if} $R$ nonempty \texttt{then} pass the token to a node in $R$ \texttt{else}
\begin{itemize}[nosep]
\item[]
\texttt{if} $s$ is different from the source 
\begin{itemize}[nosep]
\item[]
 \texttt{then} return the token to the parent and \texttt{exit}
 \end{itemize}
\begin{itemize}[nosep]
\item[]
\texttt{else exit}
\end{itemize}
\end{itemize}

\end{itemize}
\end{enumerate}

\FF

\hrule

\caption{\label{alg:Traverse-To-Broadcast}
Pseudocode executed by a node $s$ upon receiving a token while holding it.
The source initiates a token and is the first node holding the token.}
\end{figure}

When the algorithm  is invoked then nodes do not know their neighbors yet.
Each node uses the procedures \textsc{Estimate-Hidden} and \textsc{Discover} to discover the hidden neighbors in the communication graph.
If a node hears a message from a neighbor that is executing \textsc{Estimate-Hidden} with an invitation to its hidden neighbors to join in disclosing themselves and becoming children in a breadth-then-depth tree, then this is a first signal the node obtains that a broadcast has been initiated.
Upon a token's visit to a discovered neighbor, the visited node creates a list of its hidden neighbors  just to be discovered.
The token will be dispatched to visit these nodes one by one after each return.
When the list of the discovered neighbors gets exhausted, the token is returned to the parent node from which it arrived.
If the token returns to the initiating source node and all its neighbors have been already visited by the token  then the traversal terminates.


\begin{theorem}
\label{thm:traverse-to-broadcast}

Algorithm  \textsc{Traverse-To-Broadcast} accomplishes a broadcast from a single-node start in $\cO(n \log^2 N)$ rounds with high probability. 
Each node uses $\cO(\log^3 N)$ random bits.
\end{theorem}

\begin{proof} 
Each node obtains a correct upper bound on the number of hidden neighbors with high probability when executing \textsc{Estimate-Hidden}, by Lemma~\ref{lem:hidden}.
If the bound holds true then procedure \textsc{Discover} identifies the hidden neighbors correctly, by Lemma~\ref{lem:Discover}.
The token's traversal is implemented by nodes exchanging messages, and this always occurs only between neighbors in the communication graph, by Proposition~\ref{prop:hearing-only-from-neighbors}.
The token traverses the obtained breadth-then-depth spanning tree, by Fact~\ref{fact:breadth-then-depth}, and accomplishes broadcasting in the process.

Now we estimate the running time. 
Each node executes procedure \textsc{Estimate-Hidden} and \textsc{Discover} once, by the pseudocode in Figure~\ref{alg:Traverse-To-Broadcast}.
These procedures are executed sequentially, started by receiving the token for the first time.
Executing \textsc{Estimate-Hidden}  takes $\cO(\log^2 N)$ rounds, since it consists of two nested loops, by the pseudocode in Figure~\ref{alg:estimate-hidden}, each taking $\cO(\log N)$ iterations.
Therefore the total time spent on executing \textsc{Estimate-Hidden} is $\cO(n \log^2 N)$ with high probability.
A node participates only once as a hidden neighbor of a node executing \textsc{Discover}.
The sum of upper bounds on the number of hidden neighbors is at most $2^5 n$ with high probability, by Lemma~\ref{lem:hidden}.
It follows that the time spent on all the executions of \textsc{Discover} is $\cO(n \log N)$ with high probability.

Randomness is used only in procedure \textsc{Estimate-Hidden}.
Lemma~\ref{lem:hidden} gives the needed estimates on probability.
A node performs $\cO(\log^2 N)$ experiments, each requiring $\cO(\log N)$ random bits.
\end{proof}

\section{Backbone From Synchronized Start}

\label{sec:coordinated-start}

We develop a randomized algorithm to build a backbone  from a synchronized start, which means that all the nodes begin an execution together.
The algorithm runs in $\cO(\Delta\log N+\log^c N)$ time, where $c$ is a positive constant. 
The running times of algorithms \textsc{Inter$_H$} and \textsc{Intra$_H$}, associated with a backbone, are $\cO(\log N)$ and $\cO(\Delta\log N)$, respectively.


\begin{figure}[t]
\hrule

\FF

\textsf{algorithm} \textsc{Backbone-Synchronized-Start} 

\FF

\hrule

\FF

\begin{description}[nosep]
\item[\sf stage 1:] call \textsc{Find-MIS}

\item[\sf stage 2:]  call \textsc{Connect-To-MIS} 

\item[\sf stage 3:]  build an implementation of \textsc{Intra$_H$}

\item[\sf stage 4:]  build an implementation of \textsc{Inter$_H$}
\end{description}

\FF

\hrule

\caption{\label{alg:Backbone-Synchronized-Start}
Pseudocode for all nodes to start simultaneously.
Procedure \textsc{Connect-To-MIS} starts from the maximal independent set found by \textsc{Find-MIS}.}
\end{figure}

The algorithm is called \textsc{Backbone-Synchronized-Start}.
Its pseudocode  is  in Figure~\ref{alg:Backbone-Synchronized-Start}. 
An execution begins by calling two procedures.
One of them elects a maximal independent set  of nodes in the communication graph; we present it in Section~\ref{sec:finding-maximal-independent}.
The other one inter-connects the nodes in the obtained maximal independent set into a connected dominating and also connects the remaining nodes to it; this procedure is discussed in Section~\ref{sec:implementing-shortcut-connecting}.
The execution concludes with finding transmission schedules for algorithms \textsc{Inter$_H$} and \textsc{Intra$_H$} associated with the backbone; the details are given in Section~\ref{sec:implementing-local-algorithms}.

\subsection{Finding a maximal independent set}

\label{sec:finding-maximal-independent}

We use a maximal independent set as a minimal dominating set.
The procedure to find a maximal independent set is called \textsc{Find-MIS}.
Its pseudocode is given in Figure~\ref{proc:Find-MIS}.
The invoked $(N,\Theta(\log^{3} N))$-strongly-selective sequence of length $\cO(\log^{7} N)$ exists, by the fact that for each $c>0$ such that $c\leq N$, there exists an $(N,c)$-strongly-selective sequence of length $\cO(c^2\log N)$.


\begin{figure}[t]
\hrule

\FF
 
\texttt{procedure} \textsc{Find-MIS} 

\FF

\hrule

\FF

\begin{enumerate}[nosep,leftmargin=*]
\item[]
\texttt{for} $i=1$ \texttt{to} $\lg (\Delta+1)$ \texttt{do}
~ ~ ~ /$\ast$ phase $i$ $\ast$/					
\begin{enumerate}[nosep,leftmargin=*]
\item[]
\texttt{for} $j=1$ \texttt{to} $\gamma \lg N$ \texttt{do}
~ ~ ~ /$\ast$ sub-phase $j$ $\ast$/
\begin{enumerate}[nosep]
\item
Incorporate new  members:
~ ~ ~ /$\ast$ first stage of sub-phase $j$ $\ast$/
\begin{itemize}[nosep]
\item[]
\texttt{if} $s$ is neutral \texttt{then} $s$ becomes a candidate with probability $2^i/(\Delta+1)$. 
\item[]
\texttt{if} $s$ is a candidate \texttt{then} 
\begin{enumerate}[nosep]
\item[]
$s$ executes an $(N,\gamma\log^{3} N)$-strongly-selective sequence to announce its name
\end{enumerate}
\item[]
take a record of the names from the messages heard  from candidates 
\item[]
\texttt{if} $s$ is a candidate \texttt{then}
\begin{itemize}[nosep]
\item[]
\texttt{if} $s$ heard a message from some other candidate 
\begin{itemize}[nosep]
\item[]
\texttt{then} $s$ goes back to neutral 
\item[]
\texttt{else} $s$ becomes a member
\end{itemize}
\end{itemize}
\end{itemize}
\item
Convert neutral nodes to workers:
~ ~ ~ /$\ast$ second stage of sub-phase $j$ $\ast$/
\begin{itemize}[nosep]
\item[]
\texttt{if} $s$ became a member in the preceding stage \texttt{then} 
\begin{enumerate}[nosep]
\item[]
$s$ executes  an $(N,\gamma\log^{3} N)$-strongly-selective sequence to announce its name 
\end{enumerate}
\item[]
take a record of the names  heard  from new members
\item[]
\texttt{if} $s$ is neutral and $s$ has just heard some member's name \texttt{then}
\begin{itemize}[nosep]
\item[]
$s$ becomes a worker
\item[]
$s$ assigns a member with the smallest name just heard as its representative
\end{itemize}
\end{itemize}
\end{enumerate}
\end{enumerate}
\end{enumerate}

\FF

\hrule

\caption{\label{proc:Find-MIS}
Pseudocode for a node~$s$.
The invoked $(N,\Theta(\log^{3} N))$-strongly-selective sequence is a part of code and is of $\cO(\log^{7} N)$ length.
The independent set obtained as output  consists of the nodes that become members.
The constant $\gamma$ is determined by the analysis.}
\end{figure}

The procedure \textsc{Find-MIS} works in phases numbered by integers~$i$, where $1\le i\le 1+\lg\Delta$.
A phase~$i$ consists of $\gamma\lg N$ sub-phases, for a constant $\gamma>0$.
A sub-phase $j$ of phase $i$ consists of two stages: the first one serves the purpose to elect new members of a maximal independent set and the other one determines for some nodes that they will not belong to a maximal independent set under construction, such nodes will be categorized as workers.

Here is a summary of categorizations of nodes we will use.
In the beginning of an execution, all the nodes have the status of being \emph{neutral}.
This status may change so that a node becomes either a \emph{candidate}, a \emph{member} or a \emph{worker}.
We want the status of either neutral or a candidate to be temporary and  eventually a node to become either a member or a worker, such that this status, of a member or a worker, stays unchanged.

A neutral node can become a candidate in the beginning of the first stage of a phase, as determined by an outcome of a random experiment. 
By the end of the first stage, it is also determined which among the candidates graduate to members and which become neutral again.
If a candidate in a sub-phase hears in the first stage that one of its neighbors is a candidate too then it reverts to the neutral status.
A candidate in a sub-phase that does not hear that its neighbor is a candidate becomes a member.
It follows that there are no candidates at the end of a phase.
Once a node becomes a member, at the end of the first stage, then this status is permanent.
Neutral nodes that hear in the second stage of a sub-phase that some of their neighbors are members immediately become workers.
Once a node becomes a worker, at the end of the second stage, then this status is permanent.
The sets of neutral nodes, as considered in the end of sub-phases, are monotonically decreasing,  in the sense of  inclusion.

At the end of an execution, each node becomes either a member or a worker or it remains neutral.
The set of nodes that is produced as outcome, and which is to be a maximal independent one, consists of the nodes that end up as members.

Next, we argue about the correctness and efficiency of  \textsc{Find-MIS}.
This procedure operates by nodes exchanging messages, which successfully arrive to neighbors in the communication graph once transmitted by a node, by Proposition~\ref{prop:hearing-only-from-neighbors}.
We may assume that when an execution begins then phase~$0$ is has been just completed.
We begin with formulating  an invariant for a phase $i$, for $0\le i\le \lg \Delta$, which is understood to hold at the end of phase.

\begin{center}
\begin{minipage}{\pagewidth}
\textsf{Invariant for phase $i$ of procedure \textsc{Find-MIS}:} 
\FF

Each box of the pivotal grid contains at most $(\Delta+1)/2^i$ neutral nodes.
\end{minipage}
\end{center}

Observe that initially the invariant holds for $i=0$, by the properties of pivotal grid, see~Section~\ref{sec:technical-preliminaries}.
Specifically, if a node belongs to a box, then all other nodes in the box are its neighbors in the communication graph, so there can be at most $\Delta$ of them.
If the invariant holds for a phase $i$ then it holds also for phase $i+1$ with high probability, which we discuss next, starting from Lemma~\ref{lem:candidates} through Lemma~\ref{lem:MIS}.


\begin{lemma}
\label{lem:candidates}

If the invariant holds for a phase of procedure \textsc{Find-MIS}, then during a sub-phase of this phase for each neutral node~$v$ the number of neutral neighbors of node~$v$ that become candidates is at most $(a+1)\lg N$ with probability at least $1-n^{-a}$,  for sufficiently large $n$. 
\end{lemma}

\begin{proof}
In the beginning of the phase, there are at most $(\Delta+1)/2^i$ neutral nodes in each box, by the invariant.
For a sub-phase, each neutral node becomes a candidate with probability $2^i/\Delta$.
So the expected number of neutral nodes in a box that become candidates in this sub-phase is at most~$2$.
Let $v$ be a neutral node.
Its neighbors are in at most $25$ boxes, so the expected number of neutral neighbors of node~$v$ that become candidates is at most $50$.

Suppose $X$ is the number of successes in a number of independent Bernoulli trials, with  the mean number of successes equal to~$\mu$.
We use the Chernoff bound which states that $\Pr(X\ge b) \le 2^{-b}$  for  $b\ge 6\mu$.
Let $b=c\lg n$, for $a>0$.
Then we have that $\Pr(X\ge c\lg n) \le n^{-c}$, for sufficiently large $n$.
Let a random variable~$X$ be specifically the number of neighbors of $v$ that become candidates in a sub-phase.
Choosing $c=a+1$, we obtain that at least $(a+1)\lg n$ neighbors of $v$ become candidates with probability at most $n^{-a-1}$.
This applies to each node in the network with probability at most $n^{-a}$, by the union bound.
\end{proof}


\begin{lemma}
\label{lem:hearing}

For any $a>0$ there exists $\gamma>0$ such that if the invariant  holds for a phase of procedure \textsc{Find-MIS} then each  node that executes a strongly-selective sequence to announce its name in a sub-phase of the phase is heard by all of its neighbors with probability at least $1-n^{-a}$. 
\end{lemma}

\begin{proof}
For every node, the number of its neighbors that become candidates in a sub-phase is at most $(a+1)\lg N$ with probability at least $1-n^{-a}$, by Lemma~\ref{lem:candidates}.
These nodes will execute a strongly-selective sequence to announce their names during a sub-phase.
By Proposition~\ref{prop:SINR-and-boxes}, there exists $c=\Theta(((a+1)\lg N)^3)$ such that if all the candidates execute an $(N,c)$-strongly-selective sequence, then, for each such a candidate node~$v$, there is a round in which the node~$v$  is heard by all its neighbors.
It is sufficient to take $\gamma$ at least such that $c=\gamma \lg^3 N$.
\end{proof}


\begin{lemma}
\label{lem:neutral-in-box}

For any $a>0$ there exists $\gamma>0$ such that if the invariant holds for a phase $i<\lg(\Delta+1)$ of procedure \textsc{Find-MIS} and there are at least $(\Delta+1)/2^{i+1}$ neutral nodes in phase $i+1$ in a box then there are no neutral nodes in the box after phase $i+1$  with probability at least $1-n^{-a}$.
\end{lemma}

\begin{proof}
Denote $x=\frac{\Delta+1}{2^{i+1}}$.
Each node in the box decides with probability $\frac{1}{x}$ to become a candidate in phase $i+1$.
A node in the box has fewer than $25 \cdot \frac{\Delta+1}{2^i}=50 x$ neighbors in the communication graph, by the invariant.
If exactly one neutral node in the box becomes a candidate in a sub-phase while none of its neighbors choose the same then we call it a \emph{success} in the box. 
If a success in a box occurs during a sub-phase then the node that is a candidate becomes a member and its neighbors become workers, because the candidate node does not hear from another candidate and its neighbor nodes learn of its candidate status during the sub-phase with probability at least $1-n^{-2a}$, by Lemma~\ref{lem:hearing}, for a suitable $\gamma$.
The probability of a success in a sub-phase is at least
\[
x\cdot \frac{1}{x} \Bigl(1-\frac{1}{x}\Bigr)^{50x}>\frac{1}{2 e^{50}} = p
\ ,
\]
where the number $p$ is a constant.
There exists a sufficiently large $c>0$ such that if $c\lg n$ independent trials  are performed,  each with the probability $p$ of success, then all of them are failures with probability at most $n^{-2a}$.
It suffices to take $\gamma>c$.
\end{proof}

Now we are ready to give a fact that summarizes the key properties of procedure \textsc{Find-MIS}.


\begin{lemma}
\label{lem:MIS}

For any $a>0$ there is a $\gamma>0$ such that procedure \textsc{Find-MIS} works in $\cO(\log^{8} N \log\Delta)$ rounds and produces a maximal independent set with probability at least $1-n^{-a}$.
\end{lemma}

\begin{proof} 
The number of rounds the procedure is executed is determined by its pseudocode in Figure~\ref{proc:Find-MIS}.
There are $\lg(\Delta+1)$ phases, each consisting of $\gamma \lg N$ sub-phases.
A sub-phase takes $\cO(\log^{7} N)$ rounds, because of the length of the used strongly-selective sequence.
This contributes $\cO(\log^{8} N \log\Delta)$ rounds in total.

Next, we argue about the correctness.
We show first that the invariant holds for each phase with probability at least $1-n^{-a}$, for a suitable $\gamma>0$.
The argument is by induction on the phase number~$i$.
Specifically, conditional on the invariant holding for a phase, it holds for the next phase with probability at least $1-n^{-2a}$, for a suitable constant $\gamma>0$.
The base of induction occurs for the conceptual phase number~$0$.
The invariant holds for this phase since each box contains at most $\Delta+1$ nodes by the definitions of~$\Delta$ and the pivotal grid.
In the inductive step, we show that if the invariant holds for a phase $i$ then it holds for the next phase $i+1$ with probability at least $1-n^{-2a}$.
Consider a box.
If there are at most  $\frac{\Delta+1}{2^{i+1}}$  neutral vertices in this box then certainly the invariant holds for the next phase, and otherwise there are at least  $\frac{\Delta+1}{2^{i+1}}$  vertices in the box. 
If so then, by Lemma~\ref{lem:neutral-in-box}, there are no neutral vertices after phase $i+1$ with probability at least $1-n^{-2a}$, for a suitable constant $\gamma>0$.
This completes the argument by induction.
The invariant holds for each phase with a probability that is a product of all these conditional probabilities for each phase, that is with a probability at least
\[
(1-n^{-2a})^{\lg(\Delta+1)} \ge 1-\lg(\Delta+1) \cdot n^{-2a} \ge 1-n^{-a}
\ ,
\]
by the Bernoulli's inequality.

Next we argue that when an execution is completed then the set of nodes that are members makes a maximal independent set, with a suitably high probability.
The assumption is that the invariant is satisfied for each phase.
The set of members after a phase is an independent set of nodes.
This is because when two neighbors are candidates in a sub-phase then they hear each other's names in a sub-phase with a high probability, by Lemma~\ref{lem:hearing}, and so retreat to being neutral.
To show that the independent set of members is maximal independent, it is enough to demonstrate that no neutral node remains after an execution is over: this is because then each node is either a worker or a member, so there is no room for more members.
It suffices to argue that there are no neutral node in every box.
Suppose otherwise, that there remain neutral nodes in a box.
Let $i$ be the smallest integer such that when phase~$i$ occurs then the number of neutral nodes in the box is at least $\frac{\Delta+1}{2^{i+1}}$   and at most $\frac{\Delta+1}{2^{i}}$.
By Lemma~\ref{lem:neutral-in-box}, there are no neutral nodes in the box after the phase is over with a probability at least $1-n^{-3a}$, for a suitable $\gamma>0$.
Since there are at most $n$ boxes, the probability of some neutral nodes surviving all the phases is at most $n^{-2a}$, by the union bound.

We conclude that all possible unfavorable events occur with a probability that is at most $n^{-a}$, again by the union bound.
\end{proof}

\subsection{Implementing shortcut connecting}

\label{sec:implementing-shortcut-connecting}

We describe a procedure to build a connected dominating set starting with a maximal independent set by way of implementing shortcut connecting, as it is defined in Section~\ref{sec:technical-preliminaries}.
This is done by designating  connectors which together with the independent set make a connected dominating subgraph.
It is assumed that a maximal independent set is given and each node knows whether it belongs to this set or not. 
A node in such an independent set is called a \emph{member}, to be consistent with the categorization  of some nodes as members in the sense of procedure \textsc{Find-MIS}.
The algorithm is called \textsc{Connect-To-MIS} and its pseudocode is given in Figure~\ref{proc:Connect-To-MIS}.
It is structured as four consecutive parts, which we describe in detail next.


\begin{figure}[t]
\hrule

\FF
 
\texttt{procedure} \textsc{Connect-To-MIS} 

\FF

\hrule

\FF

\begin{description}[nosep]
\item[\sf part $1$:] Non-members learn their neighbors:  
\begin{itemize}[nosep]
\item[]
\texttt{if} $s$ is a member \texttt{then} it executes a $(N,c)$-strongly selective sequence
\begin{itemize}[nosep]
\item[]
\texttt{else} $s$ records the names of member neighbors
\end{itemize}
\end{itemize}
\item[\sf part $2$:] Members learn about other members of distance at most three:
\begin{itemize}[nosep]
\item[\sf sub $1$:] any two adjacent non-members exchange information about their member neighbors 
\item[\sf sub $2$:] non-members pass what they have learned in sub-part $1$ to member neighbors
\end{itemize}
\item[\sf part $3$:] Members choose connectors and notify them:
\begin{itemize}[nosep]
\item[\sf sub $1$:] connectors of distance one from their members get notified 
\item[\sf sub $2$:] connectors of distance two from their members get notified
\end{itemize}
\item[\sf part $4$:] Nodes learn the neighborhoods in the backbone:
\begin{itemize}[nosep]
\item[\tt if] $s$ belongs to the backbone \texttt{then} 
\begin{itemize}[nosep]
\item[]
$s$ executes a $(N,c)$-strongly selective sequence 
\item[]
$s$ simultaneously records the names of neighbors heard from
\end{itemize}
\end{itemize}
\end{description}

\FF

\hrule

\caption{\label{proc:Connect-To-MIS}
Pseudocode for a node $s$.
The $(N,c)$-strongly-selective sequence is of length $\cO(\log N)$, where the constant $c>0$ is determined in analysis.
Communication in sub-parts is by way of executing suitable strongly-selective sequences.}
\end{figure}

The goal of the first part is for all the non-member nodes to learn the names of their member neighbors.
The members transmit their own names using a $(N,c)$-strongly-selective sequence of length $\cO(\log N)$, for a sufficiently large constant $c>0$, to be determined in analysis.
Every node that is not a member records the names of members, as they are heard.
The $j$-th heard name heard by a non-member node is called the \emph{$j$-th member} of this non-member node.

During the second part, members learn about other members of distance at most three in the communication graph.
This is done by having each member  communicate with its neighbors and nodes of distance two, as those were learned in the first part.
Let $x=\cO(\log^{7} N)$ be the length of an $(N,\Theta(\log^{3} N))$-strongly-selective sequence, which is the same sequence as in the implementation of procedure \textsc{Find-MIS}.
This is organized as two sub-parts such that during the first sub-part two adjacent nodes that do not belong to the maximal independent set exchange information about their  neighbors that are members, and during the second sub-part each node passes this knowledge along with the information about its own  neighbors that are members to the member neighbors, of whom there are at most~$25$. 
The details are given next.

The first sub-part proceeds as follows.
First, each non-member node $v$  chooses a number~$t_v$ that is $\Theta(\Delta)$ uniformly at random from a suitable range, which is determined in analysis.
Time is partitioned into consecutive blocks of length $x$, each devoted to executing some $(N,\Theta(\log^{3} N))$-strongly-selective sequence by suitable nodes.
Additionally, we join blocks into consecutive groups of $25$ blocks each.
Then node~$v$ is active during group~$t_v$ in every round in $[25x t_v+1,\ldots,25x(t_v+1)]$, 
in the following sense: in the $j$-th block of group $t_v$, where $0\le j\le 24$.
More precisely, in rounds $[25x t_v+jx+1,\ldots,25x t_v+(j+1)x$, node~$v$ transmits its $j$-th member and its own name using an $(N,\Theta(\log^{3} N))$-strongly-selective sequence.
Each node $w$ records all the names heard in this execution.
For every name that $u$ heard, node $w$ records the round number~$f_u$ when the name of~$u$ was heard and the node~$g_u$ who sent it.

The second sub-part proceeds similarly, with the following modifications.
First, each group consists of $49$ blocks.
This is because there is at most one member in a box of the pivotal grid and $49$ is an upper bound on the number of boxes containing nodes that are two hops away from a node,  as argued in Section~\ref{sec:technical-preliminaries}.
Second, each block is associated with some known member, either in one-hop neighborhood, as learned in the first part, or in two-hop neighborhood, as learned in  the first sub-part of this second part. 
In the latter case, such member's name, say~$u$, is transmitted along with  its associated forwarding node $g_u$ and the successful round number~$f_u$.
Third, at the end of this sub-part, each member~$w$ additionally records all members' names $u$ heard in this sub-part, together with one or two node(s)  associated with this name, and one or two, respectively,  round number(s) when successful transmissions between the node(s) and the member took place. 
In case there are many  intermediate nodes or pairs thereof, an arbitrary selection of one such a configuration per a member's name is made:  for each member's name~$u$, let $g_{u,1}$ denote a one-hop neighbor and $g_{u,2}$ be two-hop neighbor associated with this name, and let $f_{u,i}$ be associated with a successful transmission round of node~$g_{u,i}$, for $i=1,2$.
In case there is only one connector associated with a member of name~$u$, we denote it by~$g_u$ and its successful round number by~$f_u$, to simplify the notation.
These nodes are designated as connectors.

In the third part, members inform connectors chosen in the second part. 
Similarly as in the second part, this is carried out in two sub-parts, with first connectors of distance one and then of distance two becoming informed.
In the first sub-part, every member node $w$ transmits according to a $(N,c)$-strongly-selective sequence of length $\cO(\log N)$,  for a sufficiently large constant $c>0$, precisely $121$ times, one after another. 
The number $121$ is an upper bound on the number of member nodes in any $3$-hop neighborhood.
In the $j$-th execution of a $(N,c)$-strongly-selective sequence, node $w$ transmits its $j$-th heard member~$u$'s name  along with its associated connector~$g_u$ and round value~$f_u$, or a pair of connectors $g_{u,1},g_{u,2}$ and round numbers $f_{u,1}$ and~$f_{u,2}$.
Upon receiving such a message containing only one connector~$g_u$ and a value~$f_u$, connector~$g_u$ records that it is a connector from member~$u$ to member~$w$, as well as the successful round~$f_u$ when forwarding took place.
Upon receiving such a message containing two connectors $g_{u,1}$ and $g_{u,2}$ and  values $f_{u,1}$ and $f_{u,2}$, connector $g_{u,1}$ records that it is a connector from a member~$u$, via node $g_{u,2}$, to a member~$w$, as well as the successful round~$f_{u,1}$ when the forwarding  from it to the member~$w$ took place.
In the second sub-part, only such connectors $g_{u,1}$ are active in relaying messages.
Similarly as in the first sub-part, they use a $(N,c)$-strongly-selective sequence of length $\cO(\log N)$, for a sufficiently large constant $c>0$, precisely $121$ times one after another.
In the $j$-th execution of the $(N,c)$-strongly-selective sequence, node~$g_{u,1}$ transmits its $j$-th heard associated connector~$g_{u,2}$ for the member~$u$, along with its associated pair of members $u$ and $w$  and round value~$f_{u,2}$. 
Upon receiving such a message, connector~$g_{u,2}$ records that it is a connector from the member~$u$ to the member~$w$, via node~$g_{u,1}$, as well as the successful round~$f_{u,2}$ when the forwarding  from it to the connector~$g_{u,1}$ took place.
A non-member node is in the backbone when it  is a connector to some pair of members.

In the final fourth part, all backbone nodes, both members and connectors,  learn their neighborhoods in the backbone.
In order to accomplish it, they transmit their own names using an $(N,c)$-strongly-selective sequence of length $\cO(\log N)$, for a sufficiently large constant $c>0$. 
Each backbone node records all the receives names as its neighbors in the backbone.


\begin{lemma}
\label{lem:connect-to-mis}

Procedure \textsc{Connect-To-MIS} builds a connected dominating subnetwork satisfying all the specifications of a backbone in~$\cO(\Delta\log^{7} N)$ time with high probability, when starting with a maximal independent set of nodes.
\end{lemma}

\begin{proof}
It is sufficient to show that the algorithm \textsc{Connect-To-MIS} implements shortcut connecting in wireless networks, which makes Propositions~\ref{pro:shortcut-connecting-general} and~\ref{pro:shortcut-connecting-communication-graphs} applicable.
This follows from the fact that an execution of \textsc{Connect-To-MIS} is a systematic enlargement of a maximal independent set of nodes according to the requirements for shortcut connecting. 
We rely on the property that nodes add edges by sending messages to neighbors  in the communication graph, which always arrive successfully, by Proposition~\ref{prop:hearing-only-from-neighbors}.
A detailed argument follows.

The first part makes each non-member learn the names of all its member neighbors.
There are at most $25$ such neighbors, since there is at most one member in a box of the pivotal grid. 
If the constant~$c$ is sufficiently large, than executing a $(N,c)$-strongly-selective sequence of length~$\cO(\log N)$ is sufficient to achieve this task, by Proposition~\ref{prop:SINR-and-boxes}.

During the second part, each member obtains either direct or relayed information from each of its one- or two-hop neighbors, about their at most $121$ member neighbors, each of distance at most three from the member.
By letting each non-member node to choose a random number in range $\Theta(\Delta)$,  for a sufficiently large range proportional to~$\Delta$, in every box of the pivotal grid, there are $\cO(\log N)$ nodes that selected the same round, with high probability.
In order to allow each of these nodes to transmit alone, an $(N,\Theta(\log^{3} N))$-strongly-selective sequence of length $x=\cO(\log^{7} N)$ can be used, by Proposition~\ref{prop:SINR-and-boxes} for $b=\cO(\log N)$ active nodes per box with high probability, by an argument similar to that in the proof of Lemma~\ref{lem:neutral-in-box}.
This applies to both sub-parts. 
Additionally,  a factor of~$25$ guarantees that in the first sub-part such situations will occur at least $25$ times,  so each non-member node will be able to successfully transmit all the names of its member neighbors. 
There might be at most $49$ members of one or two hops away from a node.
The second sub-part accomplishes its goal for similar reasons.
This results in each member learning at most~$121$ other members of at most three hops away, together with at most two  connectors.

In the third part, each member needs to send  to at most two connectors the information summarizing what it learned about other  members of distance at most three away.
To accomplish this, the members execute an $(N,c)$-strongly-selective sequence of length $\cO(\log N)$ in the first sub-part, similarly as in part one, each time for a different name of such a member, repeating for at most $121$ names.
In the second sub-part, each connector that receives such a message addressed to it, in which it occurs with another connector, relays this message to the other connector using the same procedure as in the first sub-part. 
Each node could be chosen as a connector to at most $121$ pairs of members, hence repeating the $(N,c)$-strongly-selective sequence  $121$ times, each time for a different pair of members for which it was chosen as a connector, is sufficient for relaying all such messages to all two-hop connectors.
Proposition~\ref{prop:SINR-and-boxes} is used, for a constant number~$z$ of active nodes  per box, these being only members and connectors, to guarantee a successful message exchange between the neighboring pairs of active nodes.

This is repeated one more time in part four to assure that both the members and the connectors will know about the other nodes that are either  members or connectors and that are in their range. 
The argument for correctness is similar to the one in part one, except that instead of at most one  member in each box of the pivotal grid we have a constant number of backbone nodes in a box.
\end{proof}

\subsection{Implementing local algorithms}

\label{sec:implementing-local-algorithms}

The nodes in a backbone use  local  algorithms \textsc{Intra$_H$} and \textsc{Inter$_H$}.
Both \textsc{Intra$_H$} and \textsc{Inter$_H$} are deterministic, though the latter is pre-computed by a randomized algorithm.

Algorithm \textsc{Intra$_H$} facilitates communication  among the nodes in the backbone.
It schedules all members and connectors to transmit using some $(N,c)$-strongly-selective sequence of length $\cO(\log N)$, for sufficiently large constant $c>0$.
One run is executed for preprocessing to confirm neighboring backbone nodes.
Additionally, if a non-backbone node does not have its associated backbone node yet, which is its representative node, then it selects the smallest among the heard names in the backbone as its representative. 

Algorithm \textsc{Inter$_H$} facilitates communication between the nodes not in the backbone and their representatives in the backbone.
It is specified as follows. 
We refer to the length of an $(N,\Theta(\log^{3} N))$-strongly-selective sequence by $x=\cO(\log^{7} N)$.
Each non-backbone node selects a number~$t$ in the range $[1,y]$ uniformly at random, where $y=\Theta(\Delta\log N)$, with a sufficiently large constant in front of $\Delta$, and it transmits its own name by executing an $(N,\Theta(\log^{3} N))$-strongly-selective sequence in rounds  $[t\cdot x+1,t\cdot x +x]$.
A backbone node records the names of the associated non-backbone nodes from which it receives such message as a potential representative.
After time $y\cdot x$, each backbone node arranges all its associated nodes into a list and processes it as follows. 
For each associated node on the list, it executes  some $(N,c)$-strongly-selective sequence of length $z=\cO(\log N)$, for sufficiently large constant $c>0$, transmitting the associated node's name whenever scheduled.
These executions occur  for every node on the list one after another.
The part is completed within $\Delta z$ rounds, as at most $\Delta$ associated nodes are on the list.
Each associated node~$v$ remembers its position on the list of associated nodes of its potential representative, denoted~$\sigma_v$, upon receiving a message addressed to it from such a backbone node.

Algorithm \textsc{Intra$_H$} facilitates communication among neighbors in a backbone.
It  is defined as follows: every non-backbone node $v$ transmits according to a $(N,c)$-strongly-selective sequence of length $z$ in rounds $[\sigma_v \cdot z+1,\sigma_v\cdot z +z]$.
The component $\sigma_v$ makes it possible to schedule all the workers to avoid collisions at their shared representatives, as already determined, while the $(N,c)$-strongly-selective sequence component facilitates avoiding clashes between a constant number of backbone neighbors.


\begin{lemma}
\label{lem:intra-and-inter}

The local algorithms are successfully constructed in $\cO(\Delta\log^{7} N)$ rounds with high probability.
\textsc{Intra$_H$} operates in $\cO(\log N)$ time and \textsc{Inter$_H$} operates in $\cO(\Delta\log N)$ time.
\end{lemma}

\begin{proof}
By Proposition~\ref{prop:hearing-only-from-neighbors}, messages are exchanged between neighbors in the communication graph.
Proposition~\ref{prop:SINR-and-boxes} guarantees a successful message exchange between neighboring pairs of  backbone nodes, as their number per box is constant.
Next, we argue about computing algorithms \textsc{Inter$_H$} and \textsc{Intra$_H$} associated with the backbone.

Random selections allow for all non-backbone nodes to avoid collisions at their potential representatives with high probability, while an $(N,\Theta(\log^{3} N))$-strongly-selective sequence  allows to avoid clashes between $\cO(\log N)$ non-backbone nodes of different potential representatives with high probability, since the density of backbone nodes is constant per box.
One auxiliary execution of \textsc{Inter$_H$} allows to assign by each non-backbone node its representative in the backbone, as it can hear all its neighbors in the backbone due to the fact that there is a constant number of them.

A  schedule for algorithm \textsc{Intra$_H$} can be found similarly as for part two of \textsc{Connect-To-MIS}. 
Selecting a random number and a strongly-selective sequence allows each backbone node to hear all the names of neighboring non-backbone nodes with high probability. 
Backbone nodes send acknowledgements to their non-backbone neighbors using the same $(N,c)$-strongly-selective sequence as in the schedule of \textsc{Inter$_H$}, such that all non-backbone nodes receive them.

Algorithm \textsc{Intra$_H$} makes non-backbone nodes transmit, each using an $(N,c)$-strongly-selective sequence, in the order determined by the list  created during the first part of procedure \textsc{Connect-To-MIS}, see Figure~\ref{proc:Connect-To-MIS}.
Correctness follows by the property that each execution of an $(N,c)$-strongly-selective sequence is associated with a set of pairs involving backbone and non-backbone nodes.
Hence, due to a constant density per boxes of the pivotal grid, applying the strongly-selective sequence guarantees successful deliveries of each of these pairs, by  Proposition~\ref{prop:SINR-and-boxes}.

The time this computation takes is determined by the durations of all parts.
Performance bounds of algorithms \textsc{Intra$_H$} and \textsc{Inter$_H$} follow  from their specifications.
A strongly-selective sequence is of length $\cO(\log N)$.
Algorithm \textsc{Inter$_H$} uses it no more than $\Delta$ times.
\end{proof}

\subsection{Combining ingredients into a backbone}

\label{sec:synchronized-start-combining-into-backbone}

An algorithm that builds a backbone from a synchronized start is called \textsc{Backbone-Synchronized-Start}.
Its pseudocode is in Figure~\ref{alg:Backbone-Synchronized-Start}.
The algorithm begins by executing \textsc{Find-MIS}, which produces a maximal independent set.
This is followed by \textsc{Connect-To-MIS}, which works with the maximal independent set produced by \textsc{Find-MIS}.
Finally, a backbone is completed by augmenting nodes with local algorithms, as specified in Section~\ref{sec:implementing-local-algorithms}.


\begin{theorem}
\label{thm:backbone}

Algorithm \textsc{Backbone-Synchronized-Start} builds a backbone in $\cO(\Delta\log^{7} N)$ rounds with high probability, such that \textsc{Intra$_H$} operates in $\cO(\log N)$ time and \textsc{Inter$_H$} in $\cO(\Delta\log N)$ time.
\end{theorem}

\begin{proof}
The first stages produces a maximal independent set with high probability, by Lemma~\ref{lem:MIS}.
Given a success of the first stage, the second stage constructs a connected dominating set that satisfies the specifications of a backbone, by Lemma~\ref{lem:connect-to-mis}.
The time to construct local algorithms and their performance bounds follow from Lemma~\ref{lem:intra-and-inter}.
\end{proof}

\section{Backbone From General Start}

\label{sec:backbone-general-start}

We build on algorithms presented in Sections~\ref{sec:single-node-start} and~\ref{sec:coordinated-start} to obtain a general distributed algorithm to construct a backbone.
This algorithm is called \textsc{Backbone-General-Start}.
Its pseudocode in given in Figure~\ref{alg:Backbone-General-Start}.


\begin{figure}[t]
\hrule

\FF

\textsf{algorithm} \textsc{Backbone-General-Start} 

\FF

\hrule

\FF

\begin{description}[nosep]
\item[\sf stage 1:] active nodes execute procedure \textsc{Find-MIS} 

\item[\sf stage 2:] each node in the maximal independent set executes \textsc{Traverse-To-Broadcast}\\ using a dedicated token

\item[\sf stage 3:]  all nodes execute \textsc{Backbone-Synchronized-Start} at the same round

\end{description}

\FF

\hrule

\FF

\caption{\label{alg:Backbone-General-Start}
Active nodes start from the beginning, while the other nodes join after contacted by a node visited by a token.
The surviving token imposes on all the nodes a round number to be used to start executing \textsc{Backbone-Synchronized-Start} simultaneously.}
\end{figure}

We assume that an arbitrary set of nodes is activated to start an execution.
An activated node begins by executing algorithm \textsc{Find-MIS} from Section~\ref{sec:finding-maximal-independent}. 
During an execution, if a node not originally activated receives a message, it ignores it and does not join yet. 
The outcome maximal independent set is nonempty, since the initial set $S$ of awoken nodes is non-empty.

Next, each node in the obtained maximal independent set initiates an execution of \textsc{Traverse-To-Broadcast} from Section~\ref{sec:algorithm-traverse-to-broadcast} with itself as source.
This may create multiple tokens that traverse the network concurrently.
Each token carries its source name, to which we refer as the token's \emph{name}.
The tokens compete for survival by comparing their names.  
When tokens visit nodes, they leave a name behind as a trace of visit.
A token that comes to a node formerly visited by other tokens is compared, by the name it brings, to the previous visitors.

A token carries a round number, which is interpreted as a round by which a full synchronization of the network will be accomplished, as determined by its originator.
We want nodes to maintain consistency of such synchronization  times.
This is done by following the proposal of a ``highest bidder,'' understood as a token of the maximum name seen so far.
The source nodes initialize this with their own tokens, while the others nodes wait for a token to visit.

When processing messages generated in tokens traversals, all nodes participate.
If there is a conflict between two messages needed to be sent at the same round then a message for the token with a greater source name takes precedence.
To process received tokens, a node refers to a variable \texttt{max-name}.
Once a node obtains a token, it proceeds in one of the following ways:
\begin{enumerate}
\item[(i)] if this is a first  token ever received then the node sets \texttt{max-name} to the name of the token, adopts the proposed round number  propagated by the token and proceeds to facilitate this token's further traversal by sending the required messages;

\item[(ii)] if this is a  token with a smaller name than \texttt{max-name} then this token is destroyed by the node not sending messages to facilitate the token's further traversal;

\item[(iii)] if this is a token with a  name greater than \texttt{max-name} then the node sets \texttt{max-name} to the name of the visiting token, adopts the information propagated by the token, and proceeds to facilitate this token's further traversal by sending the suitable messages.
\end{enumerate}
If there is a node that keeps a token's name and the associated information it carries then we say that the token is \emph{operative} at the node.
It is critical to have a bound on the number of tokens that are operative in nodes' neighborhoods, since each active node may contribute to collisions by generating messages needed for the corresponding token's traversal.
This is reflected in the size of strongly-selective sequences used to resolve collisions.


\begin{lemma}
\label{lem:constant-number-of-tokens}

There is a constant upper bound on the number of  tokens that are operative in a box of the pivotal grid at any round.
\end{lemma}

\begin{proof}
Tokens start their traversal originating at nodes of a maximal independent set.
There is a most one node per box in this set, since the nodes in a box induce a clique in the communication graph. 
If a token arrives at a node then the node  establishes its children in the breadth-then-depth tree.
This means that each node in the box learns about this token's arrival and if this token arrives to a child in the box while a token with a larger name has already visited the box then the smaller token gets destroyed.
A token may arrive  through a node's neighbor, and a node has neighbors in at most $25$ boxes.
If we start with at most one operative token in each box then this leads to an invariant that there are at most $25$ operative tokens in a box at all times.
\end{proof}

A token with the largest name eventually  becomes the only survivor. 
Its traversal activates the whole network and synchronizes the nodes to begin a target algorithm in the same round.
This algorithm is \textsc{Backbone-Synchronized-Start} from Section~\ref{sec:synchronized-start-combining-into-backbone},
which constructs a backbone.


\begin{theorem}
\label{thm:emulate-dfs}

Algorithm \textsc{Backbone-General-Start} creates a backbone network, for any set of initially activated nodes, in $\cO(n\log^2 N +\Delta\log^{7} N )$ rounds with high probability.
The total number of random bits per node is poly-logarithmic in $N$, and the associated local algorithms backbone algorithms operate in the same times as produced by algorithm \textsc{Backbone-Synchronized-Start}.
\end{theorem}

\begin{proof}
The algorithm resorts to the procedures that were already discussed so it suffices to revisit their performance bounds and the assumptions under which they are valid.
Procedure  \textsc{Find-MIS} works in $\cO(\log^{8} N \log\Delta)$ rounds and produces a maximal independent set with high probability, by Lemma~\ref{lem:MIS}.
Algorithm  \textsc{Traverse-To-Broadcast} accomplishes a single-source broadcast from a single-node start in $\cO(n \log^2 N)$ rounds with high probability, with each node generating $\cO(\log^3 N)$ random bits, by Theorem~\ref{thm:traverse-to-broadcast}.
These bounds remain valid with multiple tokens.
This is because of a constant upper bound on the number of  tokens operative that are operative in a box of the pivotal grid in any round, by Lemma~\ref{lem:constant-number-of-tokens}.
Algorithm \textsc{Backbone-Synchronized-Start} finds a backbone in $\cO(\Delta\log^{7} N)$ rounds with high probability, by Theorem~\ref{thm:backbone}.
These bounds added together make
$\cO(n\log^2 N +\log^{8} N \log\Delta+\Delta\log^{7} N )=\cO(n\log^2 N +\Delta\log^{7} N)$ rounds with high probability.
\end{proof}

\section{Conclusion}

We developed algorithms for broadcasting and building a backbone in the model of weak-sensitivity and weak-connectivity of wireless networks, where nodes do not know their position in a coordinate system.
This extends the related results obtained for the case when nodes  can refer to their coordinates in the plane, as obtained by Jurdzi\'nski and Kowalski~\cite{JK-DISC12} and Jurdzi\'nski et al.~\cite{JKS-ICALP-13}.

Jurdzi\'nski et al.~\cite{JKS-ICALP-13} showed a lower bound $\Omega(\min\{D\Delta,n\})$ on time performance of broadcast and wake-up in wireless networks with a single-node start. 
Combining it with the performance of algorithm for building a backbone we developed demonstrates that preprocessing a network, in order to synchronize the nodes such that they can start an execution simultaneously, decreases the expected-time performance requirements for some communication tasks in  wireless networks, for the model of weak sensitivity and weak connectivity.

Algorithms for a weak-sensitivity weak-connectivity SINR regime can be compared to those developed for a related model of (geometric) radio networks.
Radio networks allow for randomized broadcast algorithms of running time proportional to the diameter~$D$ such that other parameters contribute sub-linear factors, see Kowalski and Pelc~\cite{KowalskiP05} and Czumaj and Rytter~\cite{CzumajR06}.
Solutions for SINR networks have to efficiently combine methods of resolving collisions not only coming from simultaneously transmitting neighbors in the communication graph but also  interferences coming from other parts of the network. 
Broadcast requires time $\Omega(\min\{D\Delta\})$  in a weak-sensitivity weak-connectivity SINR model.

The concept of pivotal grid defined in terms of a coordinate system in a plane but it only serves the purpose to argue about correctness and performance bounds of algorithms but does not impact actions of nodes.
It would be interesting to develop algorithms to build a backbone with comparable performance bounds in general metric spaces, such as bounded-growth ones.

Backbones provide natural infrastructure of wireless networks.
It would be interesting to explore leveraging backbones to support dynamic communication  tasks, such as  periodic broadcast, convergecast, and routing.

\newpage

\bibliographystyle{plain}
\bibliography{barebones}

\end{document}